\pdfoutput=1
\documentclass{acta}
\usepackage{supertabular,lscape,graphicx}
\usepackage{amssymb}
\usepackage{amsmath}
\usepackage{wasysym}
\DeclareSymbolFont{ppa}{OT1}{ppl}{m}{it}
\DeclareMathSymbol{\vv}{\mathalpha}{ppa}{'166}

\newfont{\hb}{rphvb at 10pt}
\newfont{\hbo}{rphvbo at 10pt}
\newfont{\bitt}{rptmbi at 12pt}
\newfont{\bits}{rptmbi at 11pt}

\SetPages{1}{38}

\SetVol{65}{2015}

\begin{document}

\newcommand{\TabApp}[2]{\begin{center}\parbox[t]{#1}{\centerline{
  {\bf Appendix}}
  \vskip2mm
  \centerline{\small {\spaceskip 2pt plus 1pt minus 1pt T a b l e}
  \refstepcounter{table}\thetable}
  \vskip2mm
  \centerline{\footnotesize #2}}
  \vskip3mm
\end{center}}

\newcommand{\TabCapp}[2]{\begin{center}\parbox[t]{#1}{\centerline{
  \small {\spaceskip 2pt plus 1pt minus 1pt T a b l e}
  \refstepcounter{table}\thetable}
  \vskip2mm
  \centerline{\footnotesize #2}}
  \vskip3mm
\end{center}}

\newcommand{\TTabCap}[3]{\begin{center}\parbox[t]{#1}{\centerline{
  \small {\spaceskip 2pt plus 1pt minus 1pt T a b l e}
  \refstepcounter{table}\thetable}
  \vskip2mm
  \centerline{\footnotesize #2}
  \centerline{\footnotesize #3}}
  \vskip1mm
\end{center}}

\newcommand{\MakeTableApp}[4]{\begin{table}[p]\TabApp{#2}{#3}
  \begin{center} \TableFont \begin{tabular}{#1} #4 
  \end{tabular}\end{center}\end{table}}

\newcommand{\MakeTableSepp}[4]{\begin{table}[p]\TabCapp{#2}{#3}
  \begin{center} \TableFont \begin{tabular}{#1} #4 
  \end{tabular}\end{center}\end{table}}

\newcommand{\MakeTableee}[4]{\begin{table}[htb]\TabCapp{#2}{#3}
  \begin{center} \TableFont \begin{tabular}{#1} #4
  \end{tabular}\end{center}\end{table}}

\newcommand{\MakeTablee}[5]{\begin{table}[htb]\TTabCap{#2}{#3}{#4}
  \begin{center} \TableFont \begin{tabular}{#1} #5 
  \end{tabular}\end{center}\end{table}}

\newfont{\bb}{ptmbi8t at 12pt}
\newfont{\bbb}{cmbxti10}
\newfont{\bbbb}{cmbxti10 at 9pt}
\newcommand{\uprule}{\rule{0pt}{2.5ex}}
\newcommand{\douprule}{\rule[-2ex]{0pt}{4.5ex}}
\newcommand{\dorule}{\rule[-2ex]{0pt}{2ex}}
\def\thefootnote{\fnsymbol{footnote}}

\hyphenation{OGLE}

\begin{Titlepage}
\Title{OGLE-IV: Fourth Phase of the Optical Gravitational Lensing
Experiment}
\vspace*{-3pt}
\Author{A.~~U~d~a~l~s~k~i$^1$,~~
M.\,K.~~S~z~y~m~a~ñ~s~k~i$^1$~~
and~~G.~~S~z~y~m~a~ñ~s~k~i$^2$}
{$^1$Warsaw University Observatory, Al.~Ujazdowskie~4, 00-478~Warszawa, Poland\\
e-mail: (udalski,msz)@astrouw.edu.pl\\
$^2$Warsaw University of Technology, Institute of Micromechanics and
Photonics, ul.~¶w.~A.~Boboli~8, 02-525~Warszawa, Poland\\
e-mail: g.szymanski@mchtr.pw.edu.pl}
\end{Titlepage}

\Abstract{We present both the technical overview and main science drivers of the
fourth phase of the Optical Gravitational Lensing Experiment (hereafter
OGLE-IV). OGLE-IV is currently one of the largest sky variability surveys
worldwide, targeting the densest stellar regions of the sky. The survey
covers over 3000 square degrees in the sky and monitors regularly over a
billion sources.

The main targets include the inner Galactic Bulge and the Magellanic
System. Their photometry spans the range of $12<I<21$~mag and
$13<I<21.7$~mag, respectively. Supplementary shallower Galaxy Variability
Survey covers the extended Galactic bulge and 2/3 of the whole Galactic
disk within the magnitude range of $10<I<19$~mag. All OGLE-IV surveys
provide photometry with milli-magnitude accuracy at the bright end. The
cadence of observations varies from 19--60 minutes in the inner Galactic
bulge to 1--3 days in the remaining Galactic bulge fields, Magellanic
System and the Galactic disk.

OGLE-IV provides the astronomical community with a number of real time
services. The Early Warning System (EWS) contains information on two
thousand gravitational microlensing events being discovered in real time
annually, the OGLE Transient Detection System (OTDS) delivers over 200
supernovae a year. We also provide the real time photometry of
unpredictable variables such as optical counterparts to the X-ray sources
and R~Coronae Borealis stars.

Hundreds of thousands new variable stars have already been discovered
and classified by the OGLE survey. The number of new detections will be
at least doubled during the current OGLE-IV phase. The survey was
designed and optimized primarily to conduct the second generation
microlensing survey for exoplanets. It has already contributed
significantly to the increase of the discovery rate of microlensing
exoplanets and free-floating planets.}


\vspace*{-3pt}
\Section{Introduction}
The Optical Gravitational Lensing Experiment (OGLE) is a long-term
large-scale photometric sky survey focused on sky variability. Established
at the beginning of 1990s the project has continuously been providing
important discoveries in a variety of fields of modern astrophysics --
already for more than 22 years.

The OGLE sky survey began regular sky monitoring on April, 12, 1992 as one
of the first generation microlensing sky surveys. Followed Paczyñski's
(1986) ideas on the properties and detection rates of gravitational
microlensing events in our Galaxy as well as proposed by him application of
microlensing for the search for dark matter in the Galactic halo, the first
generation microlensing projects (MACHO, EROS, OGLE) started at that time
hunting for these extremely rare phenomena. These searches required a
completely new strategy of conducting observations -- long-term
(years-long) monitoring of the densest stellar regions in the sky
(Magellanic Clouds, Galactic center and disk). Fast identification of the
first microlensing events (Alcock \etal 1993 -- LMC, Udalski \etal 1993 --
Galactic bulge) encouraged further intensive observations. Nevertheless, it
took almost two decades before the final microlensing results on the dark
matter content in the halo finally ruled out low mass objects as a
significant component of the Galactic halo (Alcock \etal 2000,
Tisserand \etal 2007, Wyrzykowski \etal 2011).

The first generation microlensing surveys were in fact the first large
scale variability surveys. Huge photometric databases of millions of stars
observed for several years became a unique source of data for studying
variability of the sky. Main microlensing targets, the Magellanic Clouds
and the Galactic bulge, described often as the main astrophysical
laboratories, provided millions of interesting objects of all variability
types for which precise photometry could be retrieved as a by-product of
these programs. For the first time astronomers could analyze not only
single objects of a class but also large statistical samples of different
types of stars.

The OGLE survey itself was evolving in the last two decades. The first
phase, OGLE-I, was conducted in the years 1992--1995 at the Las Campanas
Observatory, Chile, with the 1.0-m Swope telescope equipped with a single
chip CCD camera (Udalski \etal 1992). Due to limited telescope access only
the Galactic bulge was regularly monitored. OGLE-I phase led to the
discovery of the first microlensing phenomena toward the Galactic center
(Udalski \etal 1993). Many papers were published on variable objects
discovered in the Galactic bulge as well as nearby dwarf galaxies and
globular clusters monitored as sub-projects of the main survey (\eg
Udalski \etal 1994b, Kaluzny \etal 1995, Kaluzny \etal 1996).

The first major instrumental upgrade of the OGLE survey occurred in
1996. Thanks to the generous attitude to the OGLE project by the Director,
Dr. Leonard Searle, and astronomers of the Observatories of the Carnegie
Institution for Science (owner of the Las Campanas Observatory in Chile)
and OGLE's encouraging initial results, it became possible to place a new
1.3-m telescope dedicated to the OGLE project at the Las Campanas site. The
Las Campanas Observatory is one of the best astronomical observing sites
worldwide. First ground work for the new telescope started in August
1995. The f/9.2 Ritchey-Cr{\'e}tien telescope with three-lens field
corrector providing 1.5 degree diameter field of view was manufactured by
the DFM Engineering Inc. The ``first light'' of the Warsaw telescope was
taken on the night of February 9/10, 1996.

The OGLE-II phase began in January 1997 and lasted up to December 2000
(Udalski, Kubiak and Szymañski 1997). The new 1.3-m Warsaw telescope was
equipped with a new single chip CCD camera with a very good quality CCD
detector TeX-2048. To increase monitored area of the sky the CCD detector
was operated using the drift-scan technique. This mode of operation
increased significantly the data throughput. With a round-the-year access
to the sky new targets were added to the list of OGLE regularly monitored
objects, namely the Magellanic Clouds. Generally, the OGLE survey observing
capabilities increased by a factor of 30 in the amount of collected data
compared to the OGLE-I phase. The number of regularly monitored stars
increased from two millions to a few tens of millions.

The main scientific results of the OGLE-II phase were a further increase of
the number of detected microlensing phenomena in both the Galactic center
and Magellanic Clouds and selection and analysis of huge samples of
variable stars of many different types. Also millions of non-variable stars
were cataloged and their photometry and astrometry were released in the
form of the first generation OGLE maps (\eg Udalski \etal 1998). These
data were used for many astrophysical projects -- studies of the structure
of Magellanic Clouds (Subramaniam 2004) and the Galactic center
(Rattenbury \etal 2007), studies of interstellar extinction (Udalski 2003a)
and others.

It is worth noting here that the OGLE survey, contrary to other large-scale
sky surveys, has always used filters closely resembling the standard {\it
BVI} bands and, thus, the transformation of the OGLE photometry to the
standard system was straightforward. Thanks to that the OGLE photometric
data were always very well suited to large variety of astrophysical
applications.

In the end of 1990s preparatory studies began for another major
instrumental upgrade of the OGLE survey. At that time a wide field
multichip CCD camera was designed. The new instrument was built,
assembled and tested on the Warsaw telescope in the first half of 2001.
On June 12, 2001 regular observations of the OGLE-III phase began
(Udalski 2003b). The new eight detector mosaic camera with SITe ST-002A
$2048\times 4096$ pixel CCDs belonged to the largest imagers worldwide
at that time and the OGLE survey with regular monitoring of a few
hundred million stars continued to be one of the largest sky surveys
operating in the first decade of the 21st century.

New observing capabilities during the OGLE-III phase enabled a significant
increase of possible scientific programs and allowed undertaking new
scientific challenges. Additional important factor of the OGLE-III phase
was the implementation of the Difference Image Analysis technique (Alard
2000, Alard and Lupton 1998, Wo¼niak 2000) in the routine OGLE-III data
pipeline. The new photometric software combined with excellent weather
conditions of the Las Campanas site allowed obtaining routinely
milli-magnitude (mmag) accuracy photometry, almost on the photon noise
level in the wide range of magnitudes in all stellar fields regularly
monitored by OGLE, including the densest in the Galactic center or
Magellanic Clouds. The OGLE photometry has commonly been appreciated as the
most accurate that can be obtained from the ground and very stable over
long time scales.

The most important scientific contributions of the OGLE-III phase included
pioneering works in the extrasolar planet field -- the discovery of first
exoplanets with two new photometric techniques: transits (Udalski \etal
2002) and gravitational microlensing (Bond \etal 2004), the discovery of
the largest samples of variable stars (currently the OGLE collection
contains about half a million of periodic objects of all types -- see
Soszyñski \etal 2014 and references therein), studies on the Galactic
center and Magellanic Cloud structure and many others.

In the second half of 2000s it became clear that while the OGLE-III survey
could be in operation for several years being still competitive, its
observing capabilities were highly insufficient for undertaking new
generation surveys and projects. Some of them should be a natural step
forward after OGLE-III discoveries but with OGLE-III capabilities they were
not feasible. Therefore in 2005 an idea of a new significant upgrade of
instrumental capabilities emerged ultimately leading to the new phase of
the OGLE survey -- OGLE-IV. It was realized that much larger CCD mosaic
camera filling the entire 1.5 square diameter field of view of the Warsaw
telescope combined with much faster reading of very sensitive new
generation CCD detectors should allow increasing the observing capabilities
by almost an order of magnitude.

Financing of this project was secured by the Polish Ministry of
Science grant awarded to the OGLE team in 2006. The new instrument,
built in the years 2008--2009, was equipped with 32 science grade CCDs
of $2048\times 4102$ pixel size. During the ten-months-long break
after finishing OGLE-III observations (May 4, 2009) the new camera was
assembled at Las Campanas Observatory and installed on the Warsaw
telescope. The first images were taken on September 7/8, 2009. After
some final adjustments and fixing some encountered problems regular
monitoring of the sky by the OGLE survey was resumed on the night of
March 4/5, 2010. On that night the current -- OGLE-IV -- phase began.
The final tune-up of the camera parameters was done during the major
engineering run at the end of June 2010.

The OGLE-IV 262.5 Megapixel mosaic camera belongs to the largest imagers
worldwide and the OGLE-IV survey with over a billion regularly monitored
sources belongs to the largest photometric sky surveys. While the
gravitational microlensing is still very important science driver of the
fourth OGLE phase many other scientific still unexplored areas are also
investigated. With years of experience in the densest stellar fields in the
sky and focusing on studies of changes of brightness, OGLE-IV is currently
the largest sky variability survey.

In this paper we present a detailed description of the OGLE-IV sky
survey. In the following Sections the OGLE-IV camera and new associated
hardware are described as well as software of the OGLE-IV survey (both for
driving the camera and data pipeline). Observing strategy of the OGLE-IV
project is also presented with examples of a few initial scientific
results.

\Section{OGLE-IV Survey Design}
Switching from OGLE-III hardware to new instruments designed for the
OGLE-IV phase required replacement of not only the main scientific mosaic
camera but also major changes in the auxiliary hardware used at the Warsaw
telescope from the beginning of its operation. In the following Subsections
we describe the major components of the OGLE-IV observing set-up.

\Subsection{OGLE-IV 32 CCD Detector Mosaic Camera} The 1.3-m Warsaw
telescope is a Ritchey-Cr{\'e}tien optical design telescope with three
lens field corrector. According to the design the field of view of the
telescope with high image quality has the diameter of $\approx1.5$
degree, \ie over 300~mm (12~in) diameter in the focal plane. In the
OGLE-III phase only about 20\% of this large field of view
($35\arcm\times35\arcm$; $125\times 125$~mm; $5\times5$~in) was used
with the OGLE-III mosaic camera. Thus, the natural step to increase the
OGLE observing capacity was to fill the entire focal plane of the Warsaw
telescope with light detectors and the idea of designing a new large
field CCD mosaic camera was born. It was quickly realized that to
achieve the goal the new camera would have to be by a factor of four
larger than the existing eight CCD OGLE-III mosaic. Construction of such
a big instrument would be very challenging and the instrument would
belong to the largest imagers worldwide.

After careful investigations it was decided to build the mosaic with
$2048\times4102$ pixels CCDs (CCD44-82 type) manufactured by the E2V
Technologies (UK) Limited -- the best available CCDs at that time
(2006--2007). These $15~\mu$m pixel size detectors with very high
sensitivity and enhanced red sensitivity had already been used in many
astronomical applications, had extensively been tested in many
observatories and had commonly been thought as the best scientific CCD
detectors. It was also important that these detectors could be manufactured
in large series by E2V, in relatively short time scale of several months.

\begin{figure}[ht]
\centerline{\includegraphics[width=8.8cm]{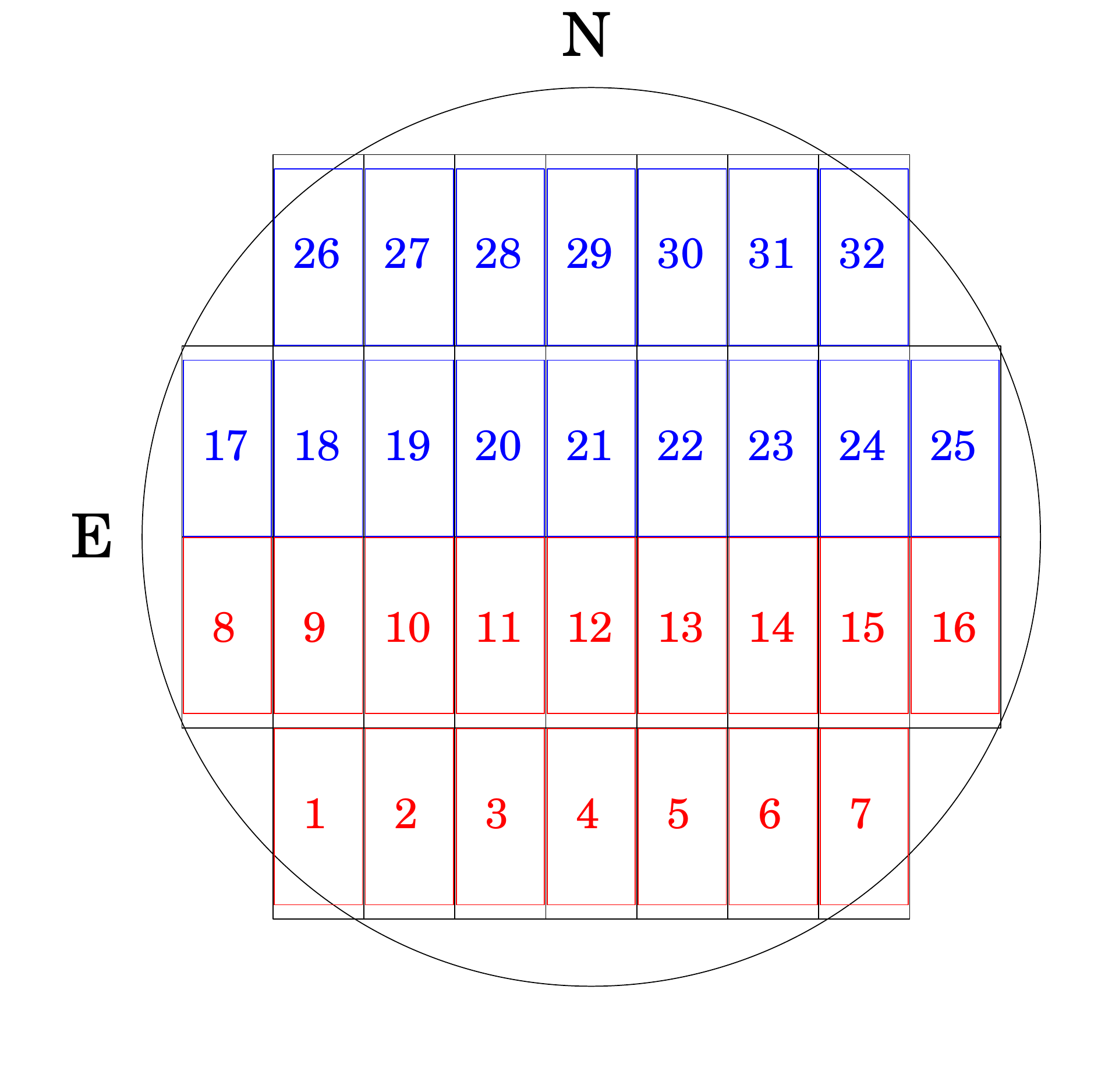}}
\vspace*{-3mm}
\FigCap{Focal plane of the Warsaw Telescope filled with the E2V 44-82
CCDs. Circle marks un-vignetted field of view of the Warsaw Telescope. N
and S indicate the directions on sky images. Red and blue marked CCDs
are driven by the controllers attached to the northern and southern sides
of the telescope (CNTL N, CNTL S), respectively.}
\end{figure}
The physical size of the imaging area of E2V CCD44-82-type CCD is $30.7
\times61.4$~mm. To fill the focal plane of the OGLE Warsaw Telescope
($\approx300$~mm diameter) we decided to use 32 detectors. Fig.~1 shows
the schematic of the focal plane of the OGLE-IV mosaic camera. 

To maximize the focal plane filling we decided to place the CCDs in four
rows with the upper two being the mirror image of the bottom two. The shape
of the field covered by the mosaic is not rectangular but rather resembling
a cross-like shape. The first and fourth rows consist of seven detectors
while the mid ones: two and three -- of nine CCDs. This design optimizes
filling of the focal plane of the Warsaw telescope. It should be noted that
the very corner detectors are somewhat vignetted in the farthest parts from
the optical axis (about 10\% of the single detector surface). However, this
effect can be successfully removed through flat-fielding. The E2V CCD44-82
detectors are three side buttable so the gaps between the neighboring CCDs
are at those sides very small, corresponding to several arcseconds on the
sky ($\approx17\arcs$ the vertical North/South gap and $\approx26\arcs$ the
middle horizontal gap). However, the fourth side of the CCD has larger dead
area of about 5~mm (electrical connections side) so the gaps between the
row 1/2 and 3/4 are larger -- of $\approx97\arcs$ on the sky.

The pixel scale of the E2V CCD44-82-type CCDs in the focal plane of
the Warsaw telescope is 0.26~arcsec/pixel. This scale is optimal for
obtaining precise photometry in dense stellar fields and excellent
seeing conditions at the Las Campanas Observatory (see Section~4).
Fig.~2 presents the distribution of the E2V CCD44-82 CCDs in the
camera's focal plane.

In the end of 2006 the financing of the upgrade to OGLE-IV phase was
granted by the Polish Ministry of Science. In the mid 2007 the contract
with the E2V company for the delivery of 32 science CCD detectors and two
additional CCDs for guiding purposes was signed. All the ordered detectors
were delivered by the fall of 2008.

The details of the new OGLE-IV camera were designed in 2007. The main goal
was to construct a relatively compact cryogenically cooled instrument.
Fig.~3 shows a general picture of the OGLE-IV camera. The main element of
the camera is the vacuum vessel where all the detectors are mounted. It
consists of four main components -- enclosure, two lids and CCD baseplate
(focal plane). All mechanical elements of the camera and new telescope
instrument plate (Section~2.2) were manufactured by Zak³ady Mechaniczne
Kazimieruk Sp.~z o.o.
\begin{figure}[t]
\centerline{\includegraphics[width=12.5cm]{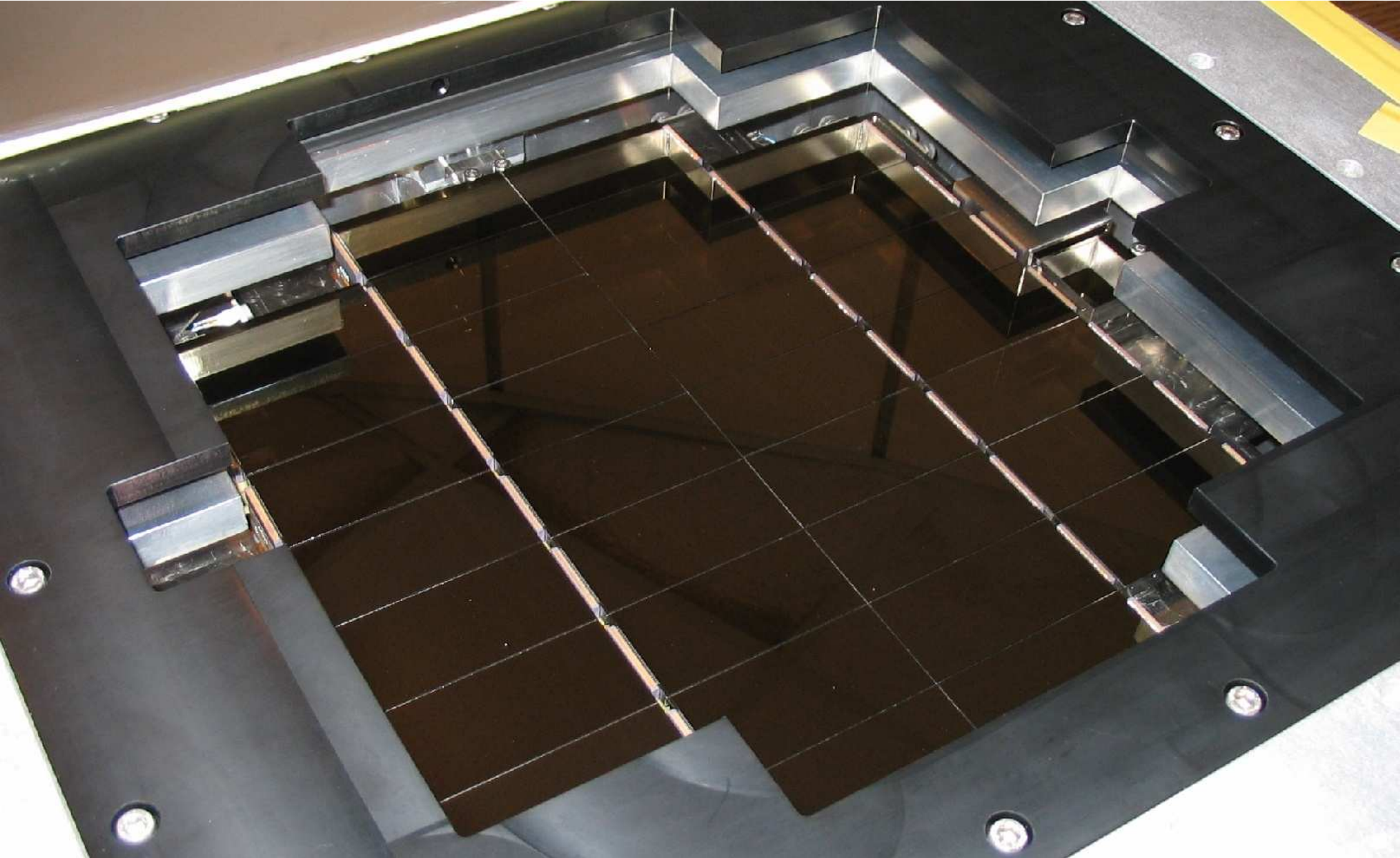}}
\vskip4pt
\FigCap{Focal plane of the OGLE-IV mosaic camera filled with the E2V
44-82 CCDs. The total size of the mosaic is about $290\times278$~mm.}
\vskip5mm
\centerline{\includegraphics[width=12.5cm]{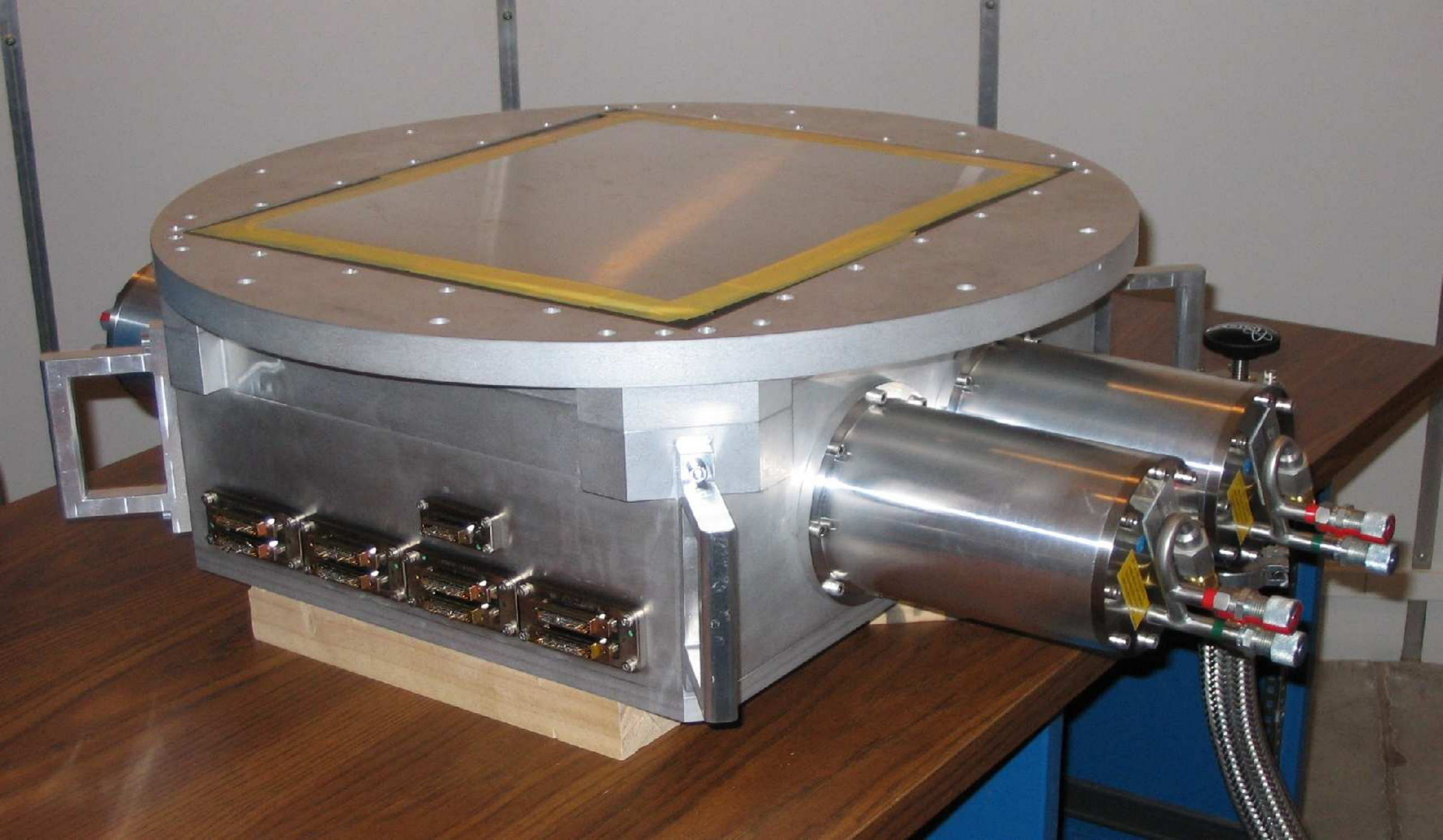}}
\vskip4pt
\FigCap{OGLE-IV mosaic camera.}
\end{figure}

\begin{figure}[ht]
\centerline{\includegraphics[width=12.5cm]{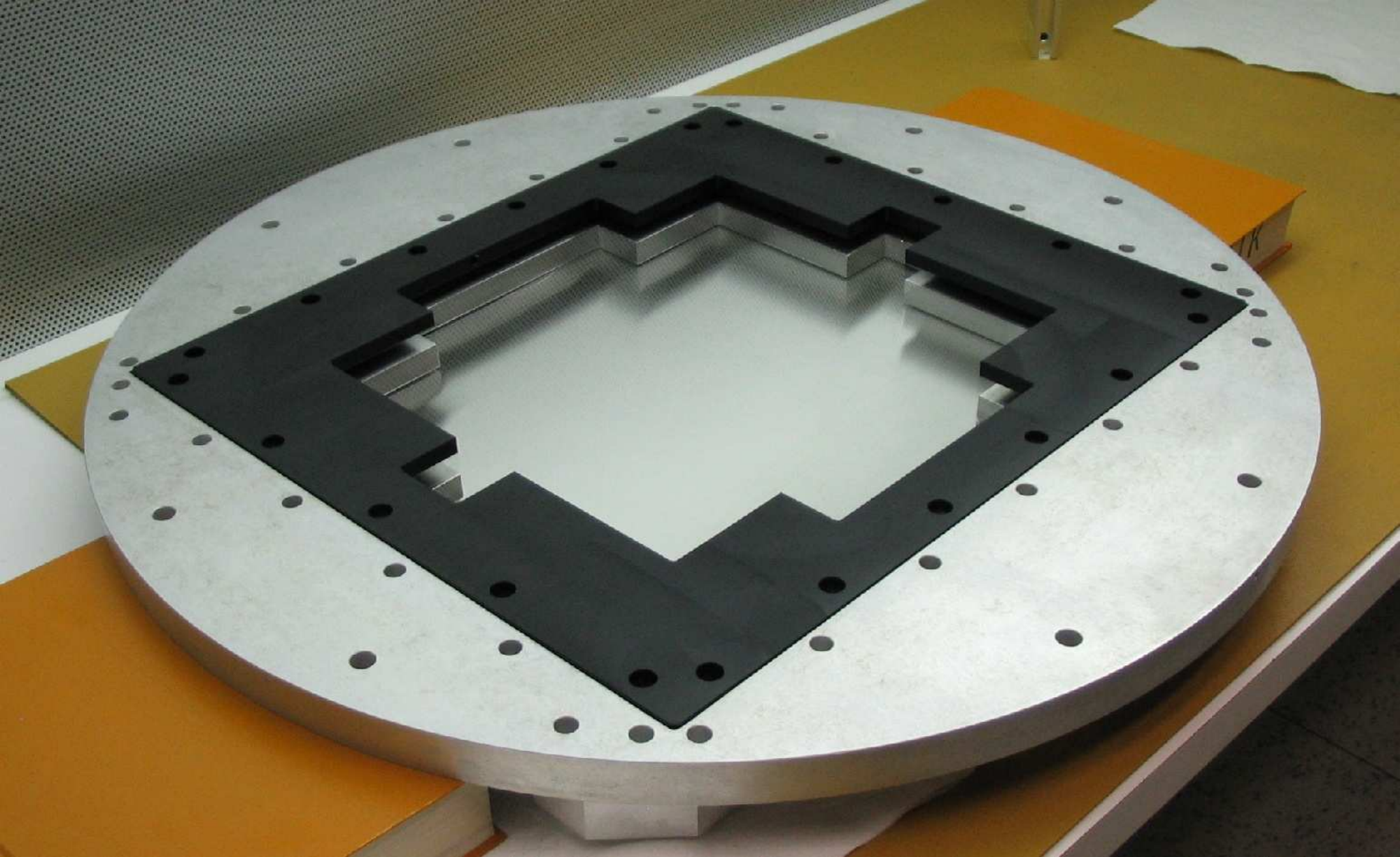}}
\vskip4pt
\FigCap{OGLE-IV mosaic camera mounting flange and window.}
\end{figure}
The size of the rectangular aluminum (6061) enclosure is $421\times
490\times 160$~mm with walls 20~mm thick. The bottom part of the vacuum
vessel is an aluminum lid: 16~mm thick aluminum plate. The top element
of the vessel is another aluminum lid being simultaneously a flange for
mounting the camera to the instrument plate of the telescope. This
element is much thicker, 50~mm, because it houses in its central part a
big window of $336\times384\times25$~mm size manufactured from the
Schott N-BK7 glass (Fig.~4). The window was coated with the multi-band
antireflective layer minimizing light losses due to reflection to less
than 1.5\% over the wide optical wavelength range (broad band
anti-reflective coating BARC-11 by ZC\&R Coatings for Optics, Inc.) Both
lids, as well as the window are sealed with viton O-rings to secure good
vacuum inside the vessel when assembled.

The main enclosure contains several ports for external devices. Four of
them are designed for mounting the cryocoolers, ten for the vacuum
connectors to secure the transmission of electrical signals to and from the
CCDs and other electrical components, one port for the vacuum gauge (979
ATV type transducer and controller by KJ Lesker Co.~Ltd) and one port for
the vacuum connections -- to the external vacuum pump (for external
prepumping) and operating non-stop ion pump (Varian Vaclon Plus 20,
StarCell) for keeping very low pressure of $10^{-6}$ mbar during the camera
operation.

The E2V CCD44-82 CCDs are packaged by the manufacturer on the $31.7\times
66.6\times 14.0$~mm invar element with customary PGA connector. The
package is designed to facilitate convenient and precise mounting of the
detector in the instrument, close packing of the detectors for mosaic
imagers and to minimize the CCD damage risk when assembling multi-detector
baseplates. Detailed design of the CCD package, information on the CCD
mounting plate design and instruction on the CCD package handling provided
by the manufacturer make proper designing of the multi-detector mosaic
instrument relatively straightforward.

\begin{figure}[ht]
\centerline{\includegraphics[width=12.5cm]{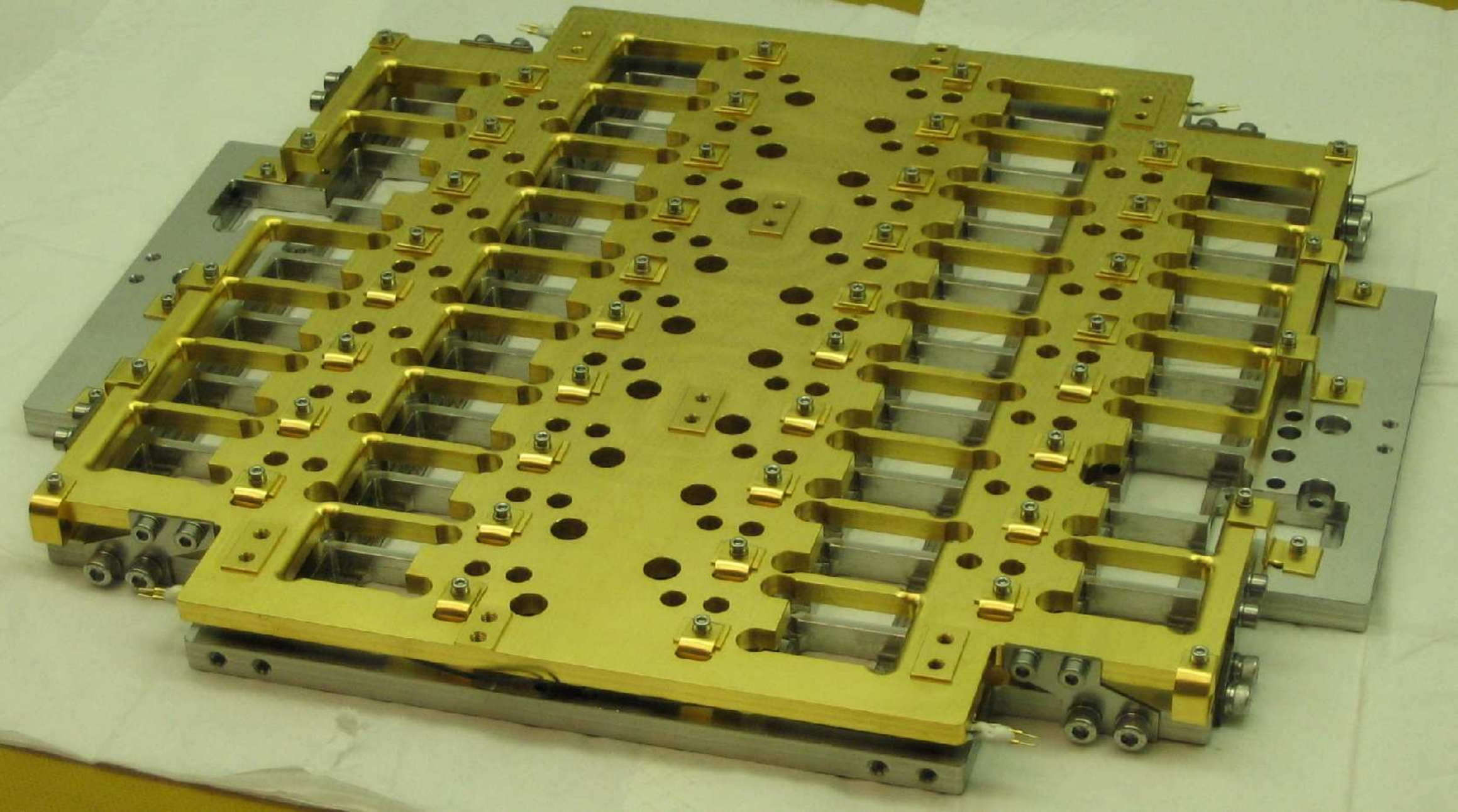}}
\vskip4pt
\FigCap{OGLE-IV mosaic camera baseplate.}
\end{figure} 
The OGLE-IV camera main baseplate (Fig.~5) consists of two components:
two plates forming a sandwich like structure -- a CCD mounting plate and
a parallel cold plate. The CCD mounting plate to which the CCDs are
directly attached is a 10~mm thick invar plate of $370\times320$~mm
size. This plate has been precisely machined to have flat (accuracy of
$5~\mu$m) mounting surface. Additionally, at each individual CCD
position a set of several holes and large rectangular opening have been
drilled and milled for mounting the CCD package and to provide access to
the PGA socket for CCD electrical connections. The second parallel cold
plate is made of copper and plated with a thin $1~\mu$m layer of gold
for lowering thermal emissivity. The cold plate is 7~mm thick and has
dimensions similar to the CCD mounting plate. Both plates are connected
with 46 gold-plated copper joints. Such a design ensures uniform removal
of heat over the entire baseplate. High thermal conductivity copper
plate acts as a cold reservoir for the CCD mounting plate. On the other
hand the invar plate has very poor thermal conductivity. However, large
number of well conducting elements joining both plates makes the heat
transfer to the CCD mounting plate very efficient and so the mounting
plate and the CCDs reach uniform temperature in a relatively short time.
Flexure of the joining elements accommodates stresses related to
different thermal expansion coefficients of the copper and invar plates.

The baseplate is mounted to the camera external enclosure with eight
$\approx110$~mm long posts made of 1.5~mm thick titanium plate, profiled
appropriately to ensure minimum of flexure and minimal heat losses. It
is cooled with four cryocoolers. Well-tested in many astronomical
applications IGC Polycold Systems Inc., Cryotiger coolers with PT-30
cooling gas were selected for the OGLE-IV mosaic camera. Each cryocooler
is mounted in the cylindrical housing attached to the appropriate port of
the main camera enclosure. The 15-m long gas lines, two for each unit,
are led through the Robotrax cable carrier system (KabelSchlepp GmbH)
and the telescope structure to the separate air-conditioned room at the
ground level of the telescope building where the compressor units are
located.

\begin{figure}[ht]
\centerline{\includegraphics[width=12.5cm]{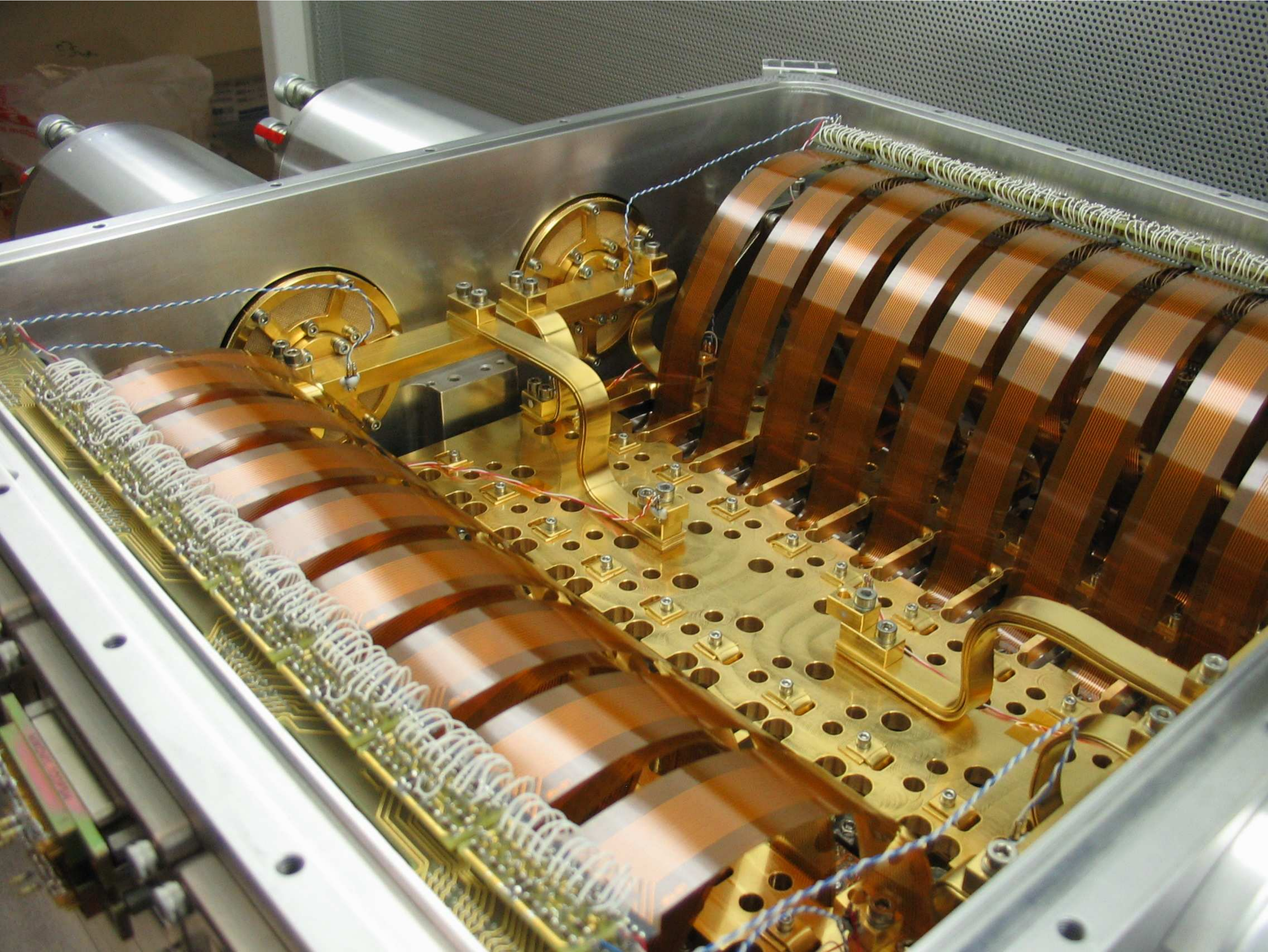}}
\vskip4pt
\FigCap{Interior of the OGLE-IV mosaic camera.}
\end{figure}
A gold plated copper container of 80~mm diameter and 12~mm depth is
mounted at the cold end of each cryocooler (Fig.~6). It contains
charcoal getter helping to maintain low vacuum for a long period of
time. The next element of the cooling system attached to the container
is a gold plated copper cold finger in the form of a bar that provides
cooling to the copper plate -- component of the baseplate -- {\it via} a
few flexible copper straps. Although the cryocoolers are basically
vibration free as they do not contain moving parts, such a solution of
the heat distribution ensures no mechanical vibrations of the baseplate.
Each cryocooler provides at least 20 Watts of cooling. They operate at
the temperature of about $-150\arcd$C at the cold finger. The number of
straps connecting the cold finger and the copper plate and their
thickness are adjusted in such a way that the CCD equilibrium
temperature is reached at about $-115\arcd$C. It can be adjusted to
higher values with the heating system mounted on the baseplate.

The electrical signals to the CCDs, temperature sensors and the internal
heating system are provided {\it via} hermetic connectors (PAVE
Technology Co., hermetic DB connectors with 200 pins, part drawing 2766)
attached to the main enclosure at appropriate ports (Fig.~3). Each of
the eight connectors for the science CCDs -- four on opposites walls of
the enclosure -- consists of two independent units. Each unit contains
two sockets connected parallelly -- one internal for connecting internal
elements of the electric signal distribution system and one external
socket for cable connection to the external hardware. Two additional
connectors (PAVE Technology Co., hermetic DB connectors with 51 pins,
part drawing 3142-1) for guider CCDs are single unit devices and also
contain internal and external sockets.

The internal electric signal distribution system (Fig.~6) was designed to
provide appropriate signals like biases, clocking voltages and to receive
video signals from each CCD detector. It consists of a series of PCB
boards, 100-pin PCB mounted plugs for connecting the system to the internal
sockets of the above-mentioned hermetic connectors and custom designed flex
cables ending with the CCD PGA plug for the connection of signals to each
of the science CCDs. Similar but much simpler system is used for single
chip guider CCDs. Additional electric signals for temperature sensors of
the camera and for the internal heating system are also distributed along
these PCBs to the appropriate connectors. Temperature sensors and heaters
distributed inside the camera are connected with thin wires to these
connectors.
\vspace*{-4pt}
\Subsection{Instrument Plate}
\vspace*{-3pt}
The instrument plate is a 18~mm thick aluminum plate mounted 224~mm from
the bottom of the telescope mirror cell with 17 aluminum and additional
four titanium posts. Such a design ensures stiffness of the construction
and minimizes flexure. The central part of the instrument plate contains
the mosaic camera mounting port. The camera is mounted to the bottom
(external plane) of the instrument plate. The room between the mirror cell
and the instrument plate is used for mounting auxiliary devices necessary
for the science operation of the OGLE-IV camera. See Fig.~7 for general
picture of this element of the OGLE-IV observing system.
\begin{figure}[p]
\centerline{\includegraphics[width=12.9cm]{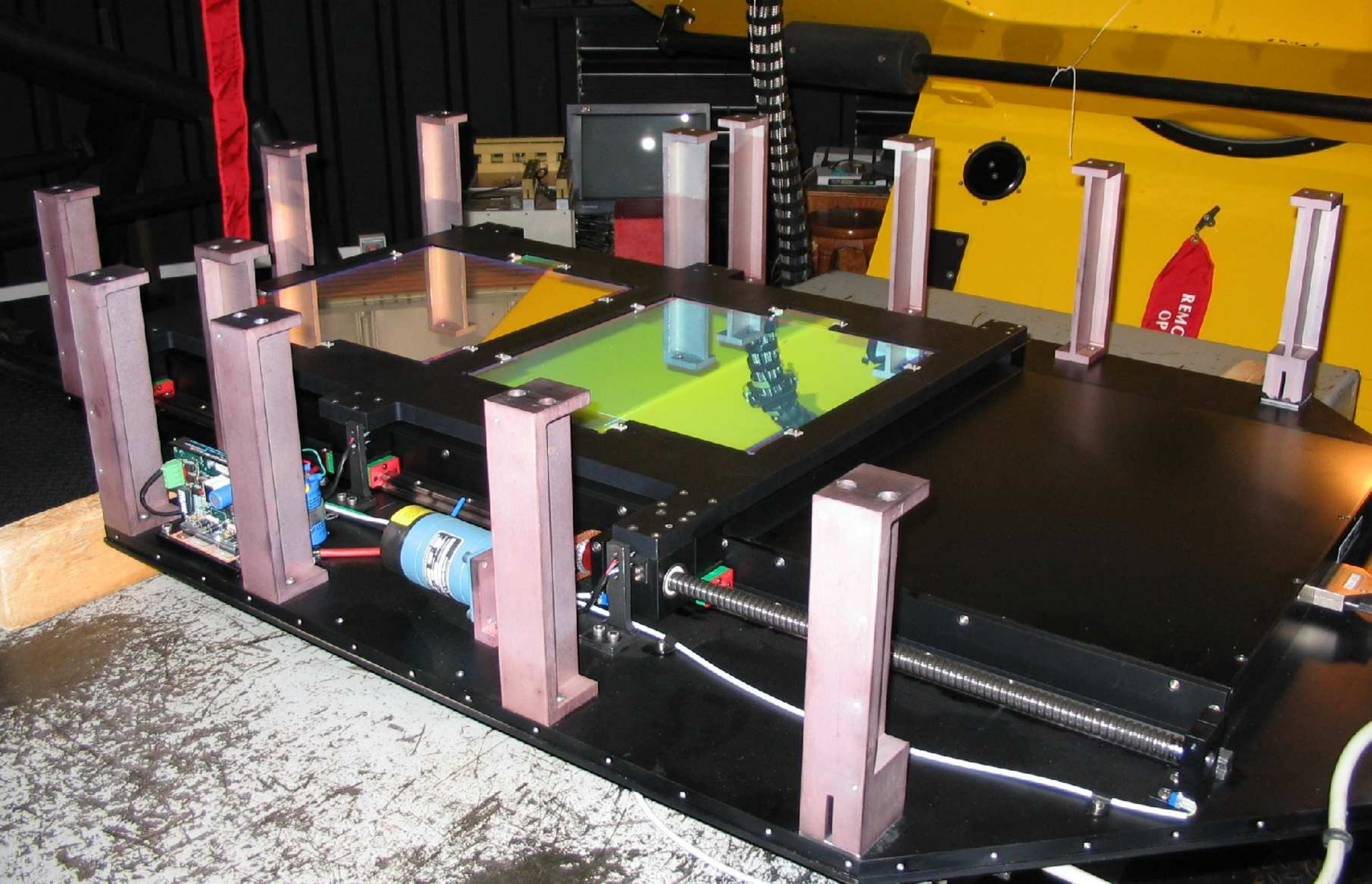}}
\vskip4pt
\FigCap{OGLE-IV instrument plate with filter holder, {\it V}- and {\it
I}-band filters, and Bonn shutter.}
\vskip7mm
\centerline{\includegraphics[width=12.9cm]{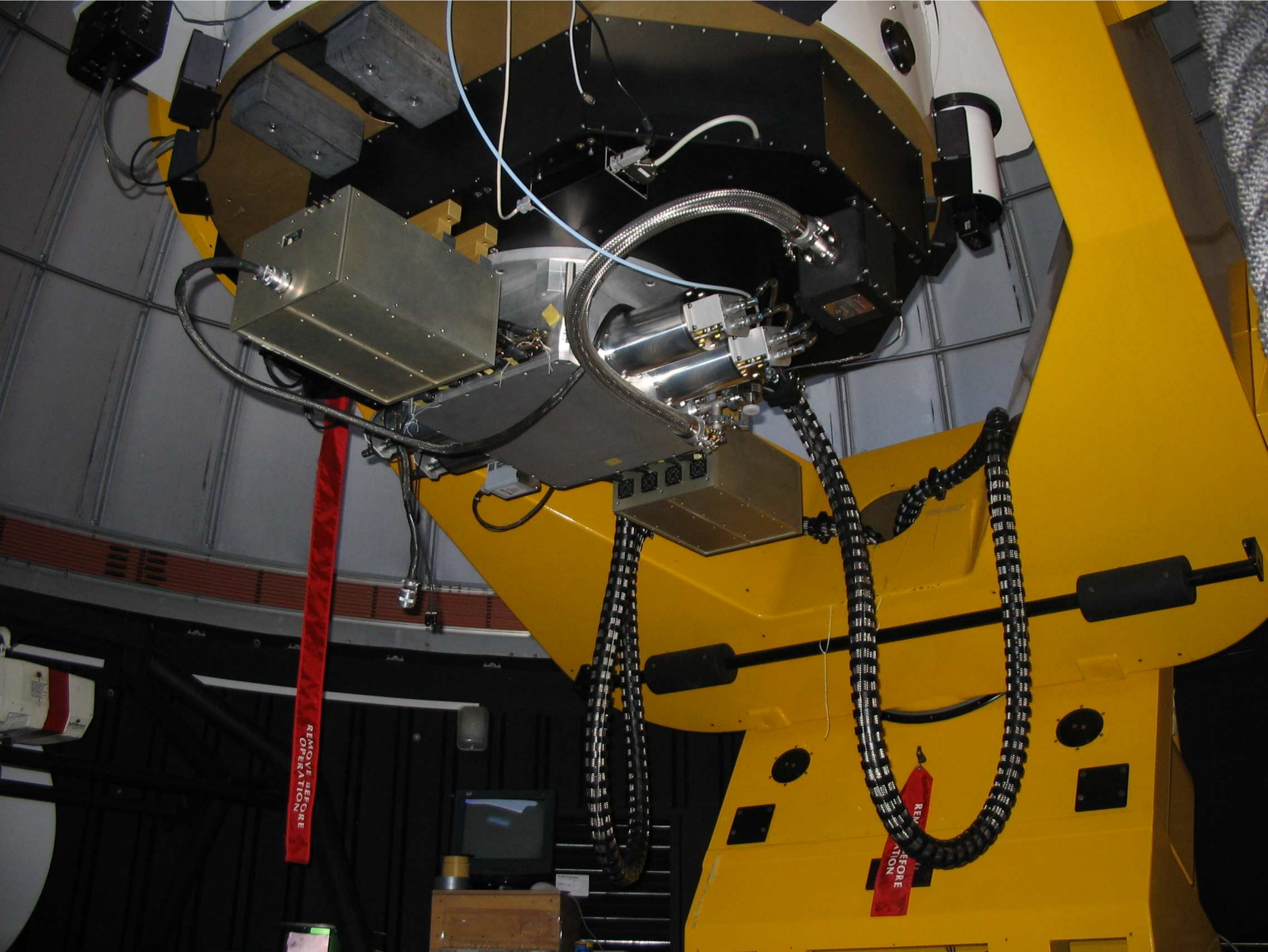}}
\vskip4pt
\FigCap{OGLE-IV mosaic camera and instrument plate on the Warsaw telescope.}
\end{figure}

The large format mechanical shutter providing $300\times 340$~mm opening
over the camera window is bolted to the internal bottom plane of the
instrumental plate. The shutter was designed and manufactured at the
Argelander Institute for Astronomy of the Bonn University, Germany ({\it
http://www.bonn-shutter.de}) and this is the OGLE suited version of
well-tested large field shutters manufactured for many photometric
surveys. The design of the shutter ensures uniform exposures, free from
the shutter error, for exposures time as short as a tenth of second. The
shutter controller is also mounted to the instrument plate, however, on
its external side.

The next important system mounted to the instrumental plate is the filter
holder. It consists of the moving filter-holding plate and supporting
hardware. Due to limited space below the telescope mirror cell only two
large size filters fitting the field of view of the Warsaw telescope are
supported. The filter set consists of a main filter and two additional
smaller filters for the guider CCDs. {\it V} and {\it I}-band interference
filters (Section~2.3) are permanently mounted in the filter holder.

The filter holder can be moved over the shutter and it is supported by two
linear guideways (HIWIN HG Series) and precisely positioned with a ball
screw driven by a computer controlled stepper motor and a set of position
sensors.

Side walls made from 1~mm thick aluminum black painted sheet close and seal
the room between the instrument plate and mirror cell protecting the
internal parts from dust and other type contamination. See Fig.~8 showing
the entire system on the Warsaw telescope.
Two CCD controllers of the OGLE-IV camera Data Acquisition System are also
mounted to the instrumental plate -- next to the OGLE-IV camera.
Additional device mounted to the instrumental plate is the ion pump,
operating permanently and ensuring very low pressure of $10^{-6}$~mbar
inside the camera.

\vskip5pt
\Subsection{Filters}
\vskip3pt
The {\it V-} and {\it I}-band interference filter sets for the OGLE-IV
camera were designed and manufactured by Asahi Spectra Co., Ltd. The
substrate on which the interference layers were coated is a 12~mm thick
quartz plate. The main component of the filter set is the science detector
filter. Its size is $310\times310$~mm. The set also contains two additional
small filters of $70\times35$~mm size with the same parameters -- for
guider CCDs. They are mounted on the sides of the same socket of the filter
holder as the main science filter.

\begin{figure}[ht]
\centerline{\includegraphics[width=13cm]{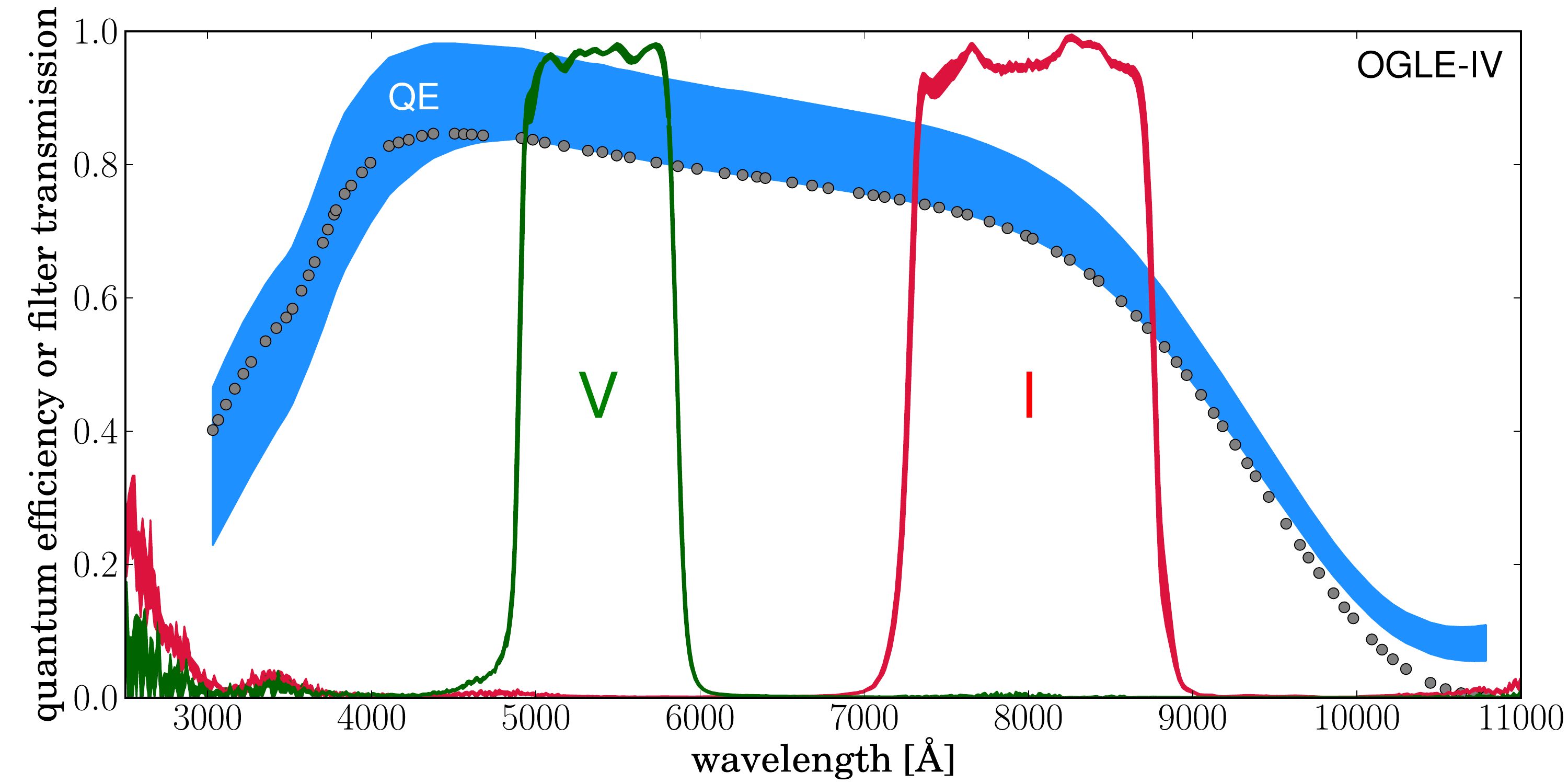}}
\vskip7pt
\FigCap{Pass-bands of the OGLE-IV filters and quantum efficiency (QE) of
the OGLE-IV CCDs. Gray dots mark QE curve from the manufacturer's
specification; blue band shows the typical QE as measured for the
actually installed CCDs; green and red curves present the
laboratory-measured filter transmission curves for {\it V}- and {\it
I}-band filters, respectively. Credit: Jan Skowron.}
\end{figure}
Fig.~9 shows the pass-band of the {\it V} and {\it I}-band OGLE filters.
Additionally, the quantum efficiency as a function of wavelength is
plotted as specified by E2V for CCD44-82 CCDs and measured for the OGLE
camera individual detectors.

While the {\it I}-band filter closely reproduces the standard
Kron-Cousins {\it I}-band filter, the transmission of the OGLE {\it
V}-band filter is somewhat different than the standard Johnson {\it
V}-band. Therefore, one can expect non-negligible color term when
transforming instrumental {\it V}-band data to the standard system (see
Section~3). Table~1 lists the average color terms, $\varepsilon_V$ and
$\varepsilon_I$, for {\it V}- and {\it I}-bands, respectively, measured
for a period of 2010--2014, for all individual CCDs of the OGLE-IV
camera. It is worth noting that the replacement of glass filters used
during the past OGLE phases with interferometric ones led to significant
increase of the performance of the OGLE-IV observing system in both {\it
V} and {\it I}-bands. The transmission of both filters is close to 100\%
-- much higher than typical transmittance of the glass filter combos.

\renewcommand{\arraystretch}{1.15}
\MakeTable{|r|@{\hspace{7pt}}c|@{\hspace{7pt}}c|@{\hspace{7pt}}c|@{\hspace{7pt}}r|}{12.5cm}{Color terms, gain and readout
noise (RN) of the OGLE-IV CCDs}
{\hline
\douprule
CCD\#   & $\varepsilon_V$ & $\varepsilon_I$ & Gain & \multicolumn{1}{c|}{RN}\\
\hline
\uprule
 1 & $-0.083\pm0.005$ & $-0.014\pm0.006$ & 1.57 & 4.90\\ 
 2 & $-0.080\pm0.001$ & $-0.010\pm0.004$ & 1.66 & 4.80\\ 
 3 & $-0.079\pm0.003$ & $-0.007\pm0.003$ & 1.78 & 5.05\\ 
 4 & $-0.081\pm0.001$ & $-0.009\pm0.004$ & 1.61 & 4.33\\ 
 5 & $-0.078\pm0.003$ & $-0.010\pm0.002$ & 1.66 & 5.50\\ 
 6 & $-0.075\pm0.003$ & $-0.008\pm0.002$ & 1.73 & 7.03\\ 
 7 & $-0.079\pm0.002$ & $-0.008\pm0.002$ & 1.78 & 9.27\\ 
 8 & $-0.082\pm0.007$ & $-0.014\pm0.005$ & 1.72 & 5.16\\ 
 9 & $-0.079\pm0.003$ & $-0.009\pm0.003$ & 1.77 & 5.33\\ 
10 & $-0.073\pm0.004$ & $-0.003\pm0.003$ & 1.75 & 5.05\\ 
11 & $-0.074\pm0.002$ & $-0.001\pm0.003$ & 1.71 & 5.18\\
12 & $-0.073\pm0.004$ & $-0.005\pm0.002$ & 1.72 & 4.88\\ 
13 & $-0.073\pm0.004$ & $-0.003\pm0.001$ & 1.65 & 4.88\\ 
14 & $-0.073\pm0.001$ & $-0.002\pm0.002$ & 1.70 & 5.42\\ 
15 & $-0.077\pm0.001$ & $-0.005\pm0.003$ & 1.74 & 4.82\\ 
16 & $-0.078\pm0.003$ & $-0.006\pm0.004$ & 1.66 & 6.38\\ 
17 & $-0.077\pm0.008$ & $-0.009\pm0.008$ & 1.64 & 5.58\\
18 & $-0.083\pm0.004$ & $-0.008\pm0.012$ & 1.86 & 8.07\\
19 & $-0.074\pm0.004$ & $-0.008\pm0.003$ & 1.73 & 6.72\\
20 & $-0.074\pm0.005$ & $-0.002\pm0.002$ & 1.69 & 4.64\\
21 & $-0.073\pm0.006$ & $-0.004\pm0.002$ & 1.65 & 4.77\\
22 & $-0.073\pm0.003$ & $-0.006\pm0.002$ & 1.66 & 4.87\\
23 & $-0.073\pm0.003$ & $-0.005\pm0.002$ & 1.80 & 5.52\\
24 & $-0.081\pm0.002$ & $-0.006\pm0.002$ & 1.72 & 6.28\\
25 & $-0.082\pm0.005$ & $-0.007\pm0.005$ & 1.73 & 4.40\\
26 & $-0.084\pm0.003$ & $-0.009\pm0.005$ & 1.64 & 9.36\\
27 & $-0.079\pm0.002$ & $-0.006\pm0.004$ & 1.75 & 11.43\\
28 & $-0.079\pm0.002$ & $-0.007\pm0.003$ & 1.83 & 5.03\\
29 & $-0.077\pm0.004$ & $-0.007\pm0.003$ & 1.68 & 4.71\\
30 & $-0.078\pm0.002$ & $-0.006\pm0.004$ & 1.78 & 4.92\\
31 & $-0.077\pm0.001$ & $-0.006\pm0.004$ & 1.74 & 4.79\\
32 & $-0.081\pm0.001$ & $-0.003\pm0.005$ & 1.88 & 5.16\\
\hline
}

\Subsection{Data Acquisition System}
\vspace*{7pt}
The electronic signals of all CCDs of the OGLE-IV mosaic camera are
fed to the CCD module of the OGLE-IV Data Acquisition System. It is
based on the Robert Leach's Astronomical Research Cameras, Inc., ARC
IV generation CCD controller system. Two identical controllers are
used in the OGLE-IV camera, each driving 16 CCD detectors: Northern
(red marked CCDs in Fig.~1): -- CCDs \#01\,--\,\#16 and Southern (blue
marked CCDs in Fig.~1): CCDs \#17\,--\,\#32.

Each controller consists of two eight-channel CCD video boards (ARC-48) for
processing the CCD signals and providing readout biases, six CCD clock
driver boards (ARC-32) providing CCD clocking voltages, one timing board
(ARC-22) generating CCD clocking signals, controlling the system and data
transfer over the fiber optic link and one utility board (ARC-50) for
additional controlling functions. All these boards are plugged to the main
data bus board providing all the necessary inter-board connections. All
boards are mounted in the 12-slot ARC system enclosure attached to the
instrument plate. The power supplies for the controllers are placed in the
telescope pedestal. Eight meter power supply cables going through the
telescope structure and Robotrax cable carriers provide all the necessary
DC voltages to the controllers.

The science CCDs work with the gain of approximately 1.7~e$^-$/ADU and
the typical readout noise (RN) is from 4 to 11~e$^-$ depending on the
detector. See Table~1 for details. The analog/digital converters of the
video chain work in the 16-bit mode, \ie ensure 65536 light levels. The
reading time of the entire array is about 18 sec.

The controllers are connected {\it via} 50-m long fiber optic cables
with PCI Interface boards (ARC-64) interfacing the ARC system with a
dedicated mosaic camera PC computer located in the server room of the
telescope control building.

The Data Acquisition System CCD software consists of a few separate
modules. Low level assembler software for clocking and reading the OGLE-IV
E2V CCD44-82 CCD detectors and controlling the ARC system is based on the
software samples and software libraries provided with the ARC
controller. It is loaded to the ARC controller when the system is
restarted.

The mosaic camera data acquisition program is run on the mosaic camera
PC computer. It starts and stops exposures, reads the mosaic CCDs and
performs many other tasks necessary for smooth operation of the CCD
array. This program can listen to commands either locally or, as during
regular operation, from external computer. During regular operation the
collected image from the OGLE-IV camera is immediately transferred by
this PC to the OGLE-IV raw images storage server {\it via} a super-fast
10 Gb Ethernet data network (LAN-CN). This transfer takes about 1.7~sec.
for a full 0.5~GByte frame what with the pixel reading time of 18~sec
makes the total reading time of the mosaic equal to about 20 sec. 

\newpage
Two guider CCDs are driven by independent guider module of the Data
Acquisition System. It is based on a significantly refurbished guiding
system used during the past phases of the OGLE survey. The guider CCDs
signal board and video processing board are somewhat modified versions of
boards used for OGLE-III mosaic camera. They are controlled by a DSP-based
controller board. This microcontroller is located at the telescope in the
so called ``red box'' mounted to the instrument plate. It controls not only
the guider CCDs but also the OGLE-IV camera filter movements and
positioning. The OGLE-IV guider CCDs are frame-transfer mode versions of
the E2V CCD44-82 CCD detectors. Storage area of the guider detector is
blocked out against incoming light by a mechanical mask mounted to the
invar plate. Thus, the guider CCDs do not need additional shutters.
However, they see the sky only during the exposure of the OGLE-IV mosaic,
when the main shutter is open.

General control of the entire guiding system (collection of guiding images,
corrections of the telescope tracking) as well as the control of the filter
(FLT) positioning and the dry nitrogen (${\rm N}_2$) anti-frost system is
done by the guider PC computer. Commands and data exchange with the ``red
box'' controller at the telescope is done {\it via} a PCI controller -- a
simplified version of the OGLE-III mosaic camera PCI controller. Guider PC
software can operate locally or, as during regular operation, it can
receive commands over the network from an external computer.

The anti-frost system of the OGLE-IV mosaic camera consists of a compressed
nitrogen cylinder located at the ground level of the telescope dome, a set
of pressure reductors, computer controlled valve and teflon tube hose to
the top lid of the camera. A small, 4~mm diameter, horizontal channel
drilled in the lid, ending over the window on one side and with the hose
fitting on another, enables the dry nitrogen flow directly over the camera
window. When the external humidity rises over 50\% the valve is opened by
the guider PC enabling slow flow of the dry nitrogen over the
camera's glass window. This prevents forming dew or frost on the large
window of the camera due to thermal emission when the camera has been
cooled down.

The main Data Acquisition System computer controlling the entire process of
data taking, {\sc ogle}, is located in the server room of the telescope
control building -- about 15~meters out of the telescope dome. This is a
2-CPUs (eight cores) SuperMicro Computers, Inc., storage server with a 24
TBytes RAID. It communicates over 1~Gb Ethernet computer network (LAN-TEL)
with the remaining Data Acquisition System computers, namely guider PC and
telescope PC. The latter computer controls operation of the 1.3-m Warsaw
telescope. It was delivered by manufacturer with the appropriate Microsoft
Windows based software. {\sc ogle} is also connected over the super-fast 10
Gb Ethernet data link (LAN-CN) with the mosaic camera PC and the computers
of the Data Analysis System. Fig.~10 presents block diagram illustrating
the main hardware of the Data Acquisition and Data Analysis Systems and
their inter-connections.
\begin{figure}[ht]
\hglue-5mm{\includegraphics[width=14cm]{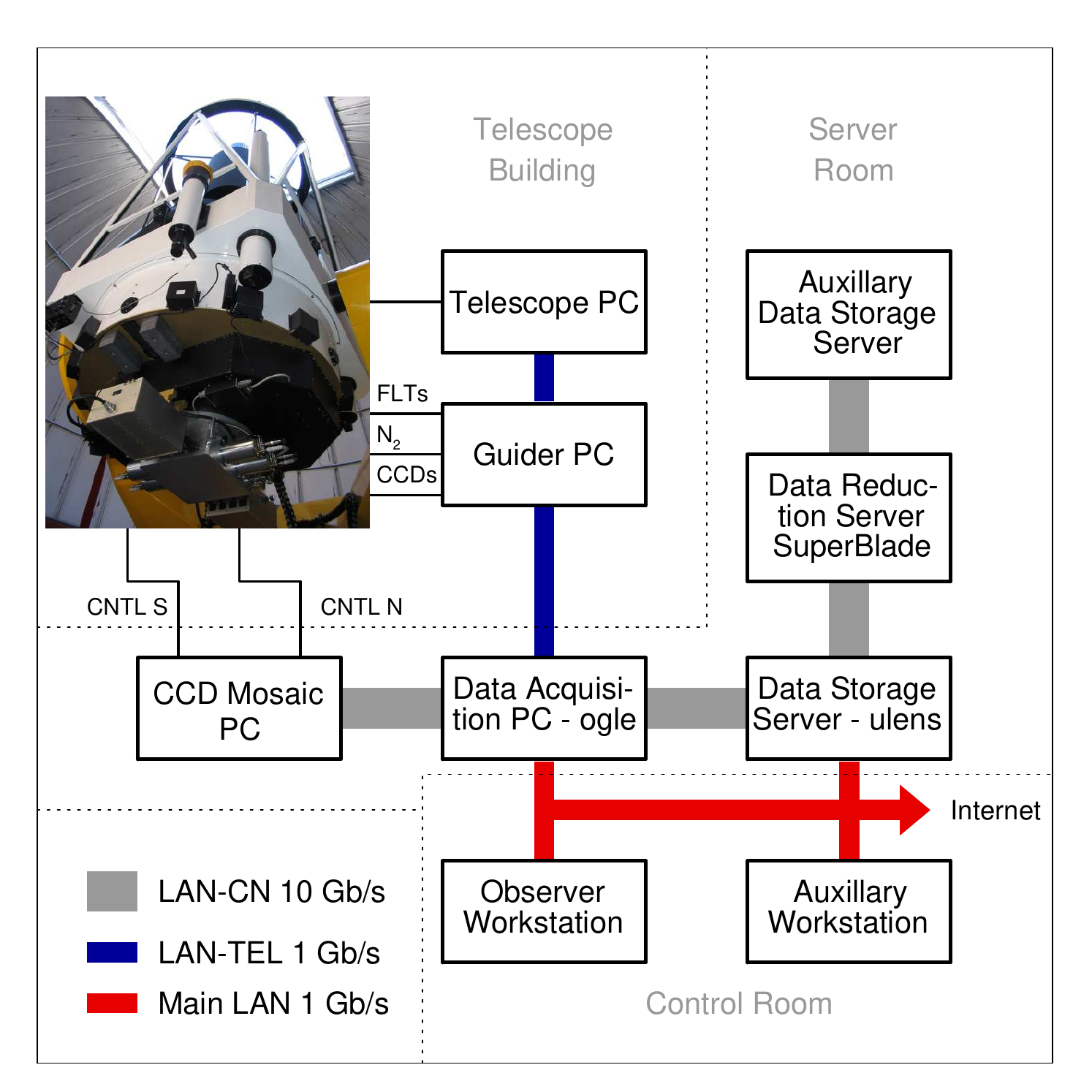}}
\FigCap{Block diagram of the OGLE-IV Data Acquisition and Data Analysis
Systems hardware.}
\end{figure}

The main Data Acquisition System software, {\sc tcs}, is run on {\sc
ogle}. This program is a graphical user interface operating under {\sc
Linux} X-windows software, using the xForms toolkit. Parallelly, it is also
a center of control -- a server exchanging appropriate commands with other
PC computers of the system. {\sc tcs} consists of three main modules
allowing user friendly, manual operation of the telescope, mosaic camera
and guider as well as of several additional utilities for effectively
conducting observations (image display, weather control, etc). {\sc tcs}
program can also operate in the batch mode. In this mode commands to the
telescope, camera or guider can be uploaded to the {\sc tcs} program in the
form of external scripts. This allows automatic operation of the entire
system. The batch mode is the main mode used during regular OGLE-IV
observations.

In the past, the {\sc tcs} program windows were displayed on the remote
{\sc ogle} monitors located in the main control room of the telescope
control building where observers conduct observations. Those monitors were
connected to the {\sc ogle} computer with a dedicated link and video signal
extending hardware. However, this solution has been replaced with a
dedicated Dell PC workstation with two large monitors running Linux
X-windows server. This workstation serves currently as the basic user
interface to the OGLE-IV Data Acquisition System and the {\sc tcs} windows
from the {\sc ogle} computer are displayed on its screens {\it via} the
X-windows protocol.

The {\sc ogle} computer RAID disk serves simultaneously as a storage area
for raw images coming from the mosaic camera PC computer. In spite of a
huge data flow of collected science images (100--200~GB uncompressed data
per night) {\sc ogle}'s RAID capacity is large enough to keep the collected
frames for several months. For data backup and transportation of the images
to the headquarters in Warsaw, the Ultrium LTO-5 tape drive (1.5/3~TB
capacity of a tape) is interfaced to the {\sc ogle} computer.

\vspace*{-9pt}
\Section{OGLE-IV Data Analysis System}
\vspace*{-5pt}
The huge number of frames collected during regular observations by the
OGLE-IV survey makes it crucial to process and analyze images in near-real
time to avoid saturation with Terabytes of data and long lags with
obtaining scientific results. The OGLE-IV Data Analysis System was designed
to process the images on-site, during the night and early morning hours.

OGLE-IV Data Analysis System contains a few components described below.
Three data pipelines are a real time software allowing image
preprocessing, main photometric reductions of collected images and fast,
almost on-the-fly photometry of selected objects. The remaining packages
are run typically when the night is over and all reductions are done.
They are designed for the detection of gravitational microlensing events
and other types of transient objects (mostly supernovae, SNe) or tracing
photometric behavior of particular variable objects.

These systems operate on instrumental OGLE-IV photometry coming from
the on-line reduction pipeline which has only been crudely calibrated
to the standard system (accuracy of the photometry zero points is
about $\pm0.15$~mag). For scientific application the OGLE-IV
photometry has to be calibrated to the standard {\it VI} system. Also
astrometry of all detected objects has to be derived.

Hardware of the OGLE-IV Data Analysis System (Fig.~10) consists of
several high performance computers. A 2-CPUs (eight cores) SuperMicro
Computers, Inc., storage server computer with 24 TBytes RAID, {\sc
ulens}, (identical to the {\sc ogle} computer of the Data Acquisition
System) has been used mainly for the image preprocessing pipeline and
storage server of prereduced images. Another Ultrium LTO-5 tape drive
is connected to {\sc ulens} for data backup. Additionally, the
selected object photometry pipeline has been implemented and run on
this computer.

In the years 2010--2014, the OGLE-IV main photometry pipeline used three
computing units from SuperMicro Computer, Inc. Every unit contained two
independent computers with two four-core CPUs each. Altogether 48 CPU cores
were available for running photometric reduction processes. These computing
capabilities ensured OGLE-IV data reductions within the 2--16 hour time
frame after the end of a night, depending on the season of year. All
computers of the Data Analysis System have been networked with the
super-fast 10 Gb Ethernet data network (LAN-CN).

In 2015 observing season these ``first generation'' computers have been
replaced by newer generation units: the {\sc ulens} computer is now a
SuperMicro dual CPU, 20 cores, 96~TB RAID storage server and the main
photometric pipeline is realized with a SuperMicro SuperBlade server.
The latter computer contains 260~CPU cores, significantly faster than
the previous CPUs and additionally 9~TB of storage area. With these
computing capabilities the OGLE-IV main photometric pipeline is now
indeed able to operate practically in real time mode -- with the
photometric reductions finished within a few minutes after a collected
frame has been preprocessed. Extended disk storage area of the OGLE-IV
Data Analysis System ensures capacities large enough for stable work
over a period of years.

The science images collected by the Data Acquisition System are stored on
the {\sc ogle} computer as raw images. The photometric prereductions
pipeline converts them into debiased and flat-fielded FITS format images,
ready for photometric reductions and stores them on the {\sc ulens} storage
server of the Data Analysis System.

The size of each full field OGLE-IV image containing all mosaic CCDs is
over 0.5~GB. Fig.~11 presents the reference image of the one of the most
dense stellar fields on the sky observed regularly by the OGLE-IV survey --
BLG505 in the Galactic bulge. It contains over six million detected stars.

To make the data handling easier, images from each CCD of the mosaic
camera (hereafter sub-fields) are stored as independent FITS files with
appropriate extensions in the file name: 01\, --\, 32 (see Fig.~1 for
correspondence of the numbers with appropriate CCD detectors). These
sub-components of the full image are then processed independently by the
OGLE-IV data reduction pipelines.
\begin{figure}[t]
\centerline{\includegraphics[width=12.9cm]{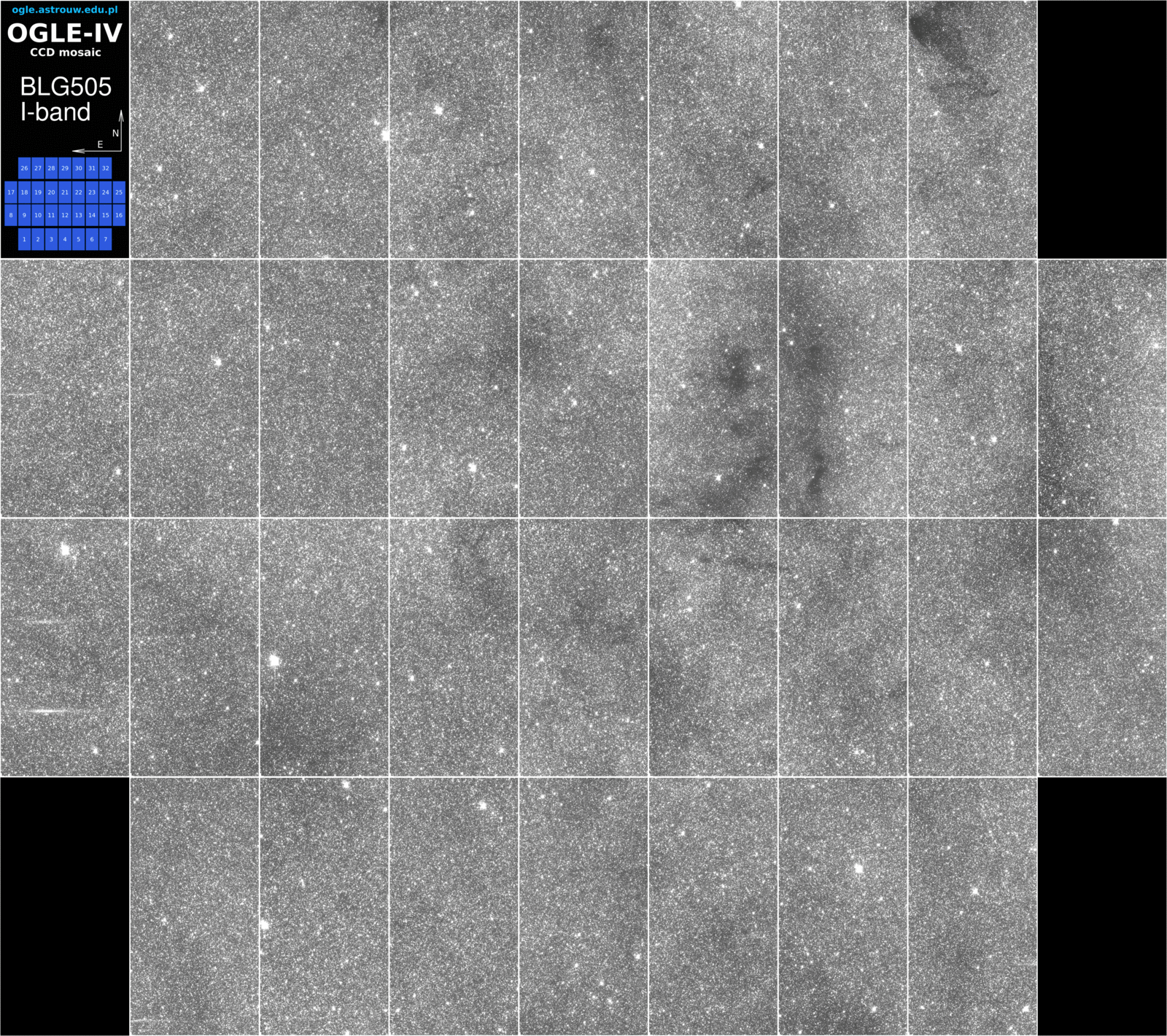}}
\vskip4pt
\FigCap{OGLE-IV mosaic camera image -- reference image of the BLG505
field (stack of five individual good quality science frames). Credit:
Szymon Koz³owski.}
\end{figure}

Every clear night a set of calibrating images: biases and sky flat-field
images are taken and new sets of master images (Zero and Flat[{\it IV}])
for each individual mosaic chip are obtained. These calibration images are
then used for de-biasing and flat-fielding of science images collected
during a night. It is worth stressing here the extremely high stability of
the OGLE-IV observing system over long periods of time. Comparison of
flat-field images taken on time scales of weeks shows negligible
variations, usually smaller than a few mmags. Therefore preparation of new
calibration images every night is not crucial for keeping the high
precision of the OGLE photometry.

The preprocessing pipeline is based on the well-tested pipeline used during
the previous phases of the OGLE survey. The main modification included
implementation of the cross-talk minimalization routines that remove
practically to zero level artifacts resulting from the cross-talk noise
generated in the multichannel video board when a bright saturated star has
been read in one of the video channels (on one of the CCDs).

Additional very important feature implemented for the OGLE-IV prereduction
pipeline is the possibility of its parallel execution what significantly
shortened the lag between image acquisition and preprocessing. During the
Galactic bulge season when the OGLE data flow is the highest, up to six
pipelines are run simultaneously ensuring practically immediate (within
5--10 minutes) preprocessing of the collected images.

The OGLE-IV main photometric data pipeline is based on the OGLE-III data
pipeline (Udalski 2003b) implemented for the OGLE-IV hardware. Some minor
modifications were introduced. The pipeline is based on the Difference
Image Analysis (DIA) technique (Alard 1998, Alard and Lupton 1999), the
Wo¼niak's (2000) implementation.

The reference images for each CCD from the OGLE-IV mosaic have the size
of $2200\times4496$ pixels. The first of the components of the reference
image (best resolution and photometric quality images), which natural
size is $2048\times4102$ pixels, is centered in this grid. It defines
the photometric zero point of the reference image and the astrometric
grid. The same first components are used to all sub-fields for a given
field, thus the same (first) image defines the basic parameters of the
reference image of a given field. Subsequent components of the reference
images which are somewhat shifted compared to the first component fill
gradually the area of the OGLE-IV reference image. The size of the
OGLE-IV reference image is large enough to fill the gaps between the CCD
detectors of the OGLE-IV mosaic camera, including two large horizontal
gaps.

The photometric reductions of each individual sub-field (individual CCD)
of the OGLE-IV field are performed on subframes to minimize the
variation of the PSF along the field (which, actually, is small) and to
simplify the data handling in dense stellar fields. The photometric
reductions of individual images are done on the grid of $4\times8$,
$2\times4$ and $1\times2$ subframes in the densest, moderately dense and
relatively empty stellar fields observed by the OGLE-IV survey,
respectively. Each subframe is reduced independently by the DIA
photometry software using standard procedures (\cf Wo¼niak 2000, Udalski
2003b) and the appropriate subframe of the reference image. When
reductions of all subframes are done, the photometry of each of them is
merged into final result files for a given image. They contain
photometry of all stellar objects detected on the reference image,
photometry of ``new'' objects (undetected as stellar on the reference
image) and information about the stellar objects from the reference
image detected on the subtracted image. The OGLE-IV real time systems
are based on the latter two files.

The OGLE-IV photometry is stored in the standard OGLE databases
(Szymañski and Udalski 1993, Udalski 2003b). They are constructed for
each sub-field (corresponding to a single CCD detector of the OGLE-IV
camera) of a given field and each pass-band. The databases are updated
after each observing night when the reductions of the collected images
are finished. Before the update, poor quality images that sometimes
accidentally enter into the reduction pipeline are filtered out to
minimize contamination of databases by poor quality data.

There are two kinds of databases for each sub-field. The main database
contains photometry of all stellar objects detected on the reference
image. The zero point of this photometry is roughly calibrated with the
accuracy of about 0.15~mag. The photometry from the main databases is
further referred to as OGLE instrumental photometry. The auxiliary database
contains photometry of ``new'' objects detected on the subtracted
images. Because the reference flux of these objects is undefined, only the
excess flux measured on the difference image of the current image is stored
in this type of databases.

There are many astrophysical situations where information on the current
photometric behavior of interesting objects is crucial for their optimal
photometric coverage. During the years 2010--2014 the delays of the OGLE
standard reduction pipeline were in many cases unacceptably long.
Therefore the independent photometric pipeline was implemented. Its
primary aim was fast determination of photometry of interesting objects
selected for fast monitoring. These are typically promising microlensing
events, SNe or other transients. Technically, the reductions were
conducted in the identical way as in the main OGLE pipeline, using
standard OGLE reference images but just on the subframes containing
these objects. The photometry was available and ready for inspection to
the observer and Warsaw headquarters within a few minutes after
pre-reductions of a new image were finished. Since 2015, the new
SuperBlade data reduction server has been ensuring the real time and
lag-free photometry of the entire frame. Therefore, the pipeline
dedicated to these interesting transient objects has been modified to
avoid doubling of the reductions.

Roughly calibrated OGLE instrumental databases are used directly for all
variability studies of the OGLE targets. However, the final results require
calibration to the standard {\it VI} photometric system. Photometric
calibrations have been done in a few steps. First, the color term
coefficients of all CCD detectors of the OGLE-IV mosaic camera have been
derived based on observations of thousands non-variable stars with large
spread in color -- secondary photometric standards from the OGLE-III
photometric maps (Udalski \etal 2008b, Szymañski \etal 2011). As already
mentioned in Section~2.3, the color term coefficients: $\varepsilon_V$ and
$\varepsilon_I$, for the {\it V} and {\it I}-bands, respectively, for all
OGLE-IV camera CCDs are listed in Table~1. The {\it I}-band filter used in
OGLE-IV resembles very closely the standard {\it I}-band and the color
terms of individual detectors are very small. The difference between the
OGLE-IV {\it V}-band filter and the standard one is larger, nevertheless
the color term coefficients are still acceptable in this case and ensure
precise calibration to the standard system. On the other hand the
throughput of both OGLE-IV interference filters is much larger than the
throughput of the standard glass filters.

In the second step for each night considered to be photometric the zero
points of the standard {\it VI} photometry were derived based on
observations of the OGLE-IV calibrating fields, namely fields fully
overlapping with OGLE-III photometric maps containing thousands of
secondary photometric standards. For each subframe of the sub-field (the
entity on which the basic DIA photometric reductions are performed) the
average aperture correction was derived allowing conversion of the OGLE
DIA instrumental photometry to the absolute signal scale. Then the
comparison of the standard OGLE-III maps photometry with the latter,
after correcting for the color term of individual standard stars,
yielded the measurement of the photometric zero point of a given night.
Averaging several ({\it V}-band) to hundreds ({\it I}-band) such
measurements provided the mean photometric zero points for a given
night. Scatter of these measurements turned out to be an useful
indicator of the photometric stability of a night. The nights with large
scatter exceeding 0.05~mag were assumed to be non-photometric.
 
Finally, all the OGLE-IV fields observed during a photometric night were
calibrated to the standard {\it VI} system based on the zero point of the
standard photometry derived in the previous step. For each subframe of a
sub-field of a given OGLE-IV frame the aperture correction was derived
between the instrumental (OGLE-IV database) photometry and the absolute
signal scale. Then, the difference between the aperture correction and the
zero point of the standard photometry of a given night provided, after
small correction for atmospheric extinction, the measurement of the average
shift between the instrumental OGLE-IV database zero point (rough
calibration assumed during the database initialization) and the standard
photometric scale.

Final shift of the zero points between instrumental (OGLE-IV database)
and calibrated photometry is the average of many such measurements
obtained during tens/hundreds of photometric nights after removing
0.05~mag and higher outliers. The calibration of the instrumental
OGLE-IV databases to the standard system is given by the following set
of equations:
\begin{eqnarray}
(V-I) & = &  \mu \cdot [(\vv_{\rm DB} + \Delta {\rm ZP}_V) - (i_{\rm DB} + \Delta {\rm ZP}_I)] \nonumber\\
I     & = & (i_{\rm DB}   + \Delta {\rm ZP}_I) + \varepsilon_I \cdot (V-I)\\
V     & = & (\vv_{\rm DB} + \Delta {\rm ZP}_V) + \varepsilon_V \cdot (V-I) \nonumber
\end{eqnarray}
where $\Delta {\rm ZP}_V$ and $\Delta {\rm ZP}_I$ denote the shift
between the OGLE-IV instrumental and the standard photometric scale zero
points and $\mu=(1-\varepsilon_V+\varepsilon_I)^{-1}$. 

\begin{figure}[ht]
\centerline{\includegraphics[width=12cm]{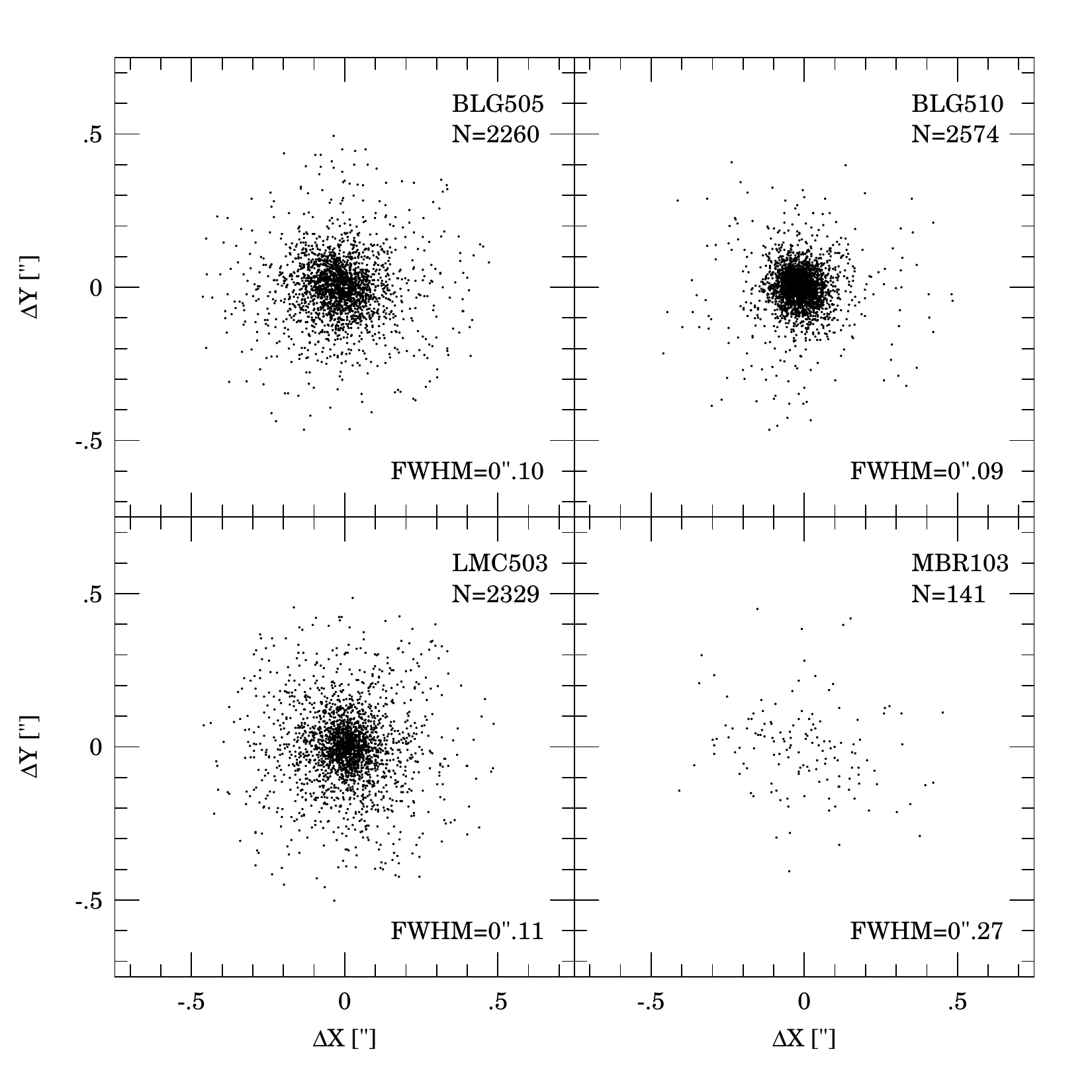}}
\vskip-3mm
\FigCap{Residuals of astrometric transformation for transformation
stars. {\it Top Panels:} Galactic bulge -- extremely dense stellar field BLG505
and typical BLG510. {\it Lower Panels:} Magellanic System -- very dense
field LMC503, very low stellar density field MBR103.}
\end{figure}
Astrometry of the OGLE-IV fields has been derived in similar way as during
the OGLE-III phase (Udalski \etal 2008a). For each of the sub-fields
(individual CCD detectors) of the OGLE-IV targets the bright stars from the
OGLE-IV databases were matched with the list of stars from the 2MASS Point
Source Catalog (Skrutskie \etal 2006) and then the third degree polynomial
transformations between the image $(X,Y)$ pixel grid and RA/DEC of the
2MASS coordinate system were derived. Fig.~12 shows the residuals of
transformation for transformation stars. $(X,Y)$ coordinate system
corresponds to E-W and S-N directions in the sky. Top panels present
residuals for the most dense Galactic bulge field BLG505 and typical field
of somewhat lower stellar density -- BLG510. The lower panels show
residuals of transformation for the typical dense stellar field in the
Magellanic System (LMC503) and typical field of very low stellar density
there (MBR103).

Generally, the accuracy of astrometric transformations is from about
0\zdot\arcs1 to 0\zdot\arcs3 in empty fields -- checked additionally
on the overlapping parts of OGLE-IV images. The systematic errors of
the 2MASS RA/DEC coordinate system are of the order of 0\zdot\arcs1
(Cutri
\etal 2000).
\vskip7pt
\Subsection{Real Time Services}
\vskip3pt
The features of the OGLE sky survey which are particularly appreciated by
the astronomical community are the real time services providing photometry
of selected types of variable objects in real time. Such data can
facilitate follow up, both ground and space observations of interesting
types of variables, for example transients or variable stars undergoing
rapid, unpredictable large brightness variations.

The Early Warning System (EWS, Udalski \etal 1994a) has been
historically the first such system. It has been providing information on
the on-going gravitational microlensing events and has been the base of
the microlensing science field for 20 years, enabling, for example,
formation of several microlensing follow-up groups like PLANET (Cassan
\etal 2012), $\mu$FUN (Gould \etal 2010) and other. During the following
OGLE phases the EWS system was re-implemented, increasing significantly
the rate of microlensing detections with enhanced observing capabilities
of the OGLE survey (Udalski 2003b).

Implementation of the EWS for the OGLE-IV survey was one of the high
priority tasks after starting the OGLE-IV phase. The first observing season
(2010) was devoted to collecting good quality images, hence enabling real
time reductions of the OGLE-IV targets. Starting from 2011, the OGLE-IV EWS
system was back in operation. The number of the Galactic bulge fields --
the main targets for microlensing research -- covered by the EWS increased
gradually from the most microlensing-efficient central Galactic bulge ones
in 2011 to all regularly observed Galactic center microlensing fields in
2013 (100 fields in 2013, reduced to 85 from 2014 on). The number of
microlensing detections reached the rate of about 2000 per year -- the
increase by a factor of three compared to the OGLE-III phase.

Two other real time systems implemented during the OGLE-III phase --
XROM and RCOM (Udalski 2008) have also been reimplemented for OGLE-IV.
These systems provide real time photometric monitoring of the optical
counterparts of X-ray sources (XROM) and R Coronae Borealis stars
(RCOM). Photometric behavior of these stars is unpredictable. OGLE-IV
XROM and RCOM systems provide astronomical community with a unique
possibility of tracing current photometric behavior of these very
interesting variable stars. The OGLE survey also continues regular
monitoring of the Einstein Cross gravitational lens (Huchra's lens).
OGLE unique coverage and precise photometry of its all four images
extend now for over 17 years (Udalski \etal 2006). OGLE-IV real time
data system for the Einstein Cross will be reimplemented in the next
months.

In 2012 a new OGLE-IV real time system was implemented -- OGLE-IV
Transient Detection System (OTDS, Koz³owski \etal 2013, Wyrzykowski
\etal 2014). Large area of the sky of about 650 square degrees around
the Magellanic System (the Magellanic Clouds and Bridge) regularly
monitored by the OGLE-IV survey provides excellent observational data
for the search for supernovae (SNe) or other type transient objects
like, for example, novae or dwarf novae. The SN detection rate of the
OGLE-IV survey reaches 200 objects per year placing the OGLE-IV SN
survey among the most efficient surveys of this kind worldwide. It is
worth noting that the OGLE-IV SN survey is limited to the Magellanic
System visibility period from Las Campanas. Thus, it is typically active
for about eight months per year.

\renewcommand{\arraystretch}{1.1}
\vskip7pt
\Section{OGLE-IV in Operation}

The OGLE-IV survey began regular observations of the sky on the night of
March 4/5, 2010. Since then observations have been conducted on practically
all clear nights. Typically about 330 nights per year are useful, at least
in part, for photometric observations at Las Campanas Observatory.

Fig.~13 presents the OGLE-IV sky coverage in the Galactic coordinates up to
the 2014 observing season. The shaded region marks the sky unavailable for
observations from the Las Campanas Observatory. Dark contours represent
OGLE-IV fields regularly observed while light gray -- fields observed in
the past for limited period of time.

Details of the OGLE-IV sky coverage can be found on the OGLE WWW page {\it
http://ogle.astrouw.edu.pl} under the ``Sky Coverage'' entry which contains
detailed maps of the OGLE-IV fields and OGLE-IV reference images (available
by clicking

\begin{landscape}
\begin{figure}[p]
\centerline{\includegraphics[width=21cm]{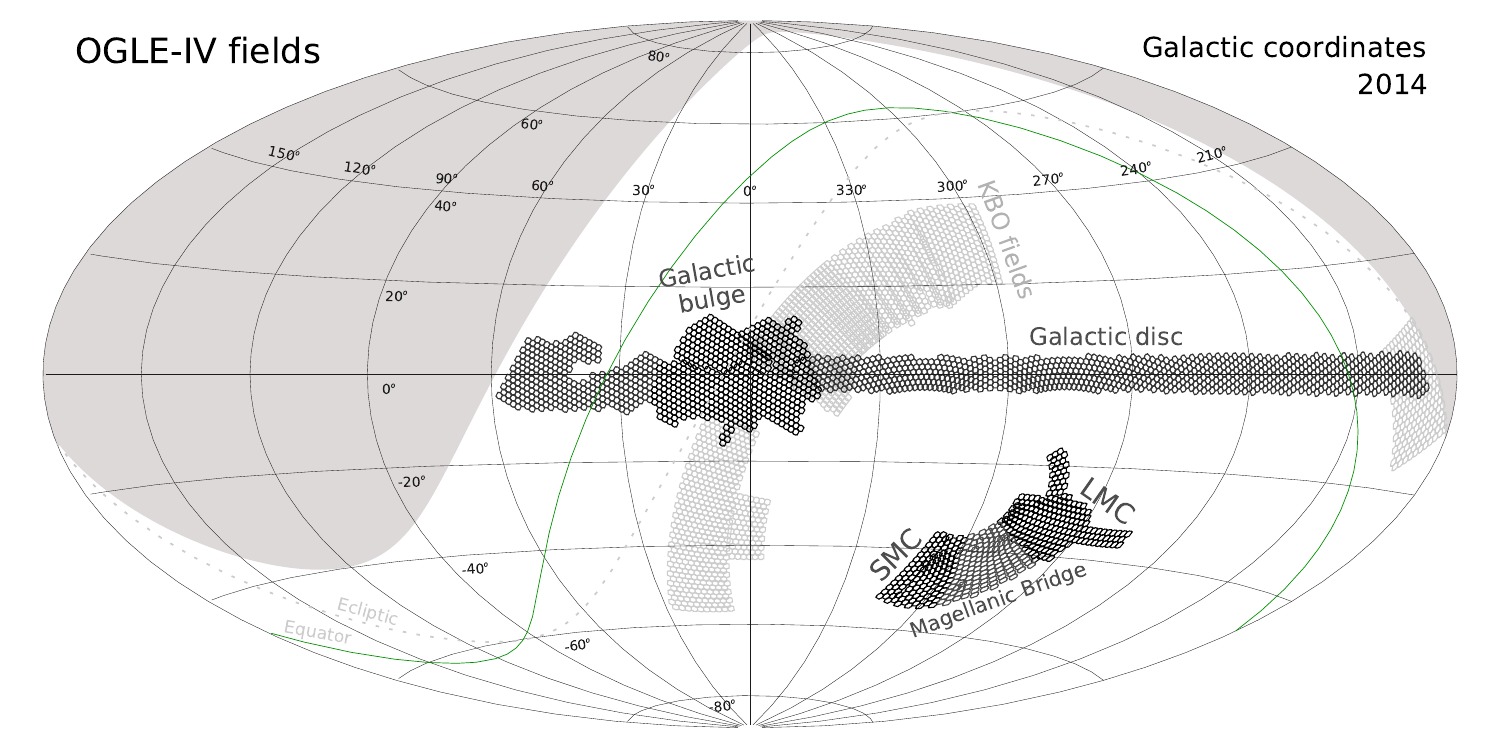}}
\FigCap{OGLE-IV sky coverage in the Galactic coordinates. Credit: Jan
Skowron.}
\end{figure}
\end{landscape}
\noindent
the field on the maps or from separate pages). We also provide
the OGLE Field Finder facility, which allows fast checking if a requested
pointing (RA/DEC) is covered by the OGLE fields.

During the 2010--2014 period of operation, the OGLE-IV survey collected
altogether over 301\,000 full science images -- 164 TeraBytes of raw
data. Table~2 lists statistics of the collected images for individual OGLE
targets described in detail in the following Subsections.

\renewcommand{\arraystretch}{1}
\MakeTable{|l|c|}{5.7cm}{Inventory of the OGLE-IV science
images collected in the years 2010--2014}
{\hline
\douprule
Object & Number of images~~\\
\hline
\uprule
Total          &  301\,528\\ 		
Galactic bulge &  113\,340\\
LMC            & ~~55\,411\\
SMC            & ~~33\,615\\
MBR            & ~~28\,432\\
Galactic disk  & ~~61\,036\\
OCKS           & ~~~5\,108\\
Other          & ~~~4\,586\\
\hline
}

\begin{figure}[ht]
\centerline{\includegraphics[width=9.5cm]{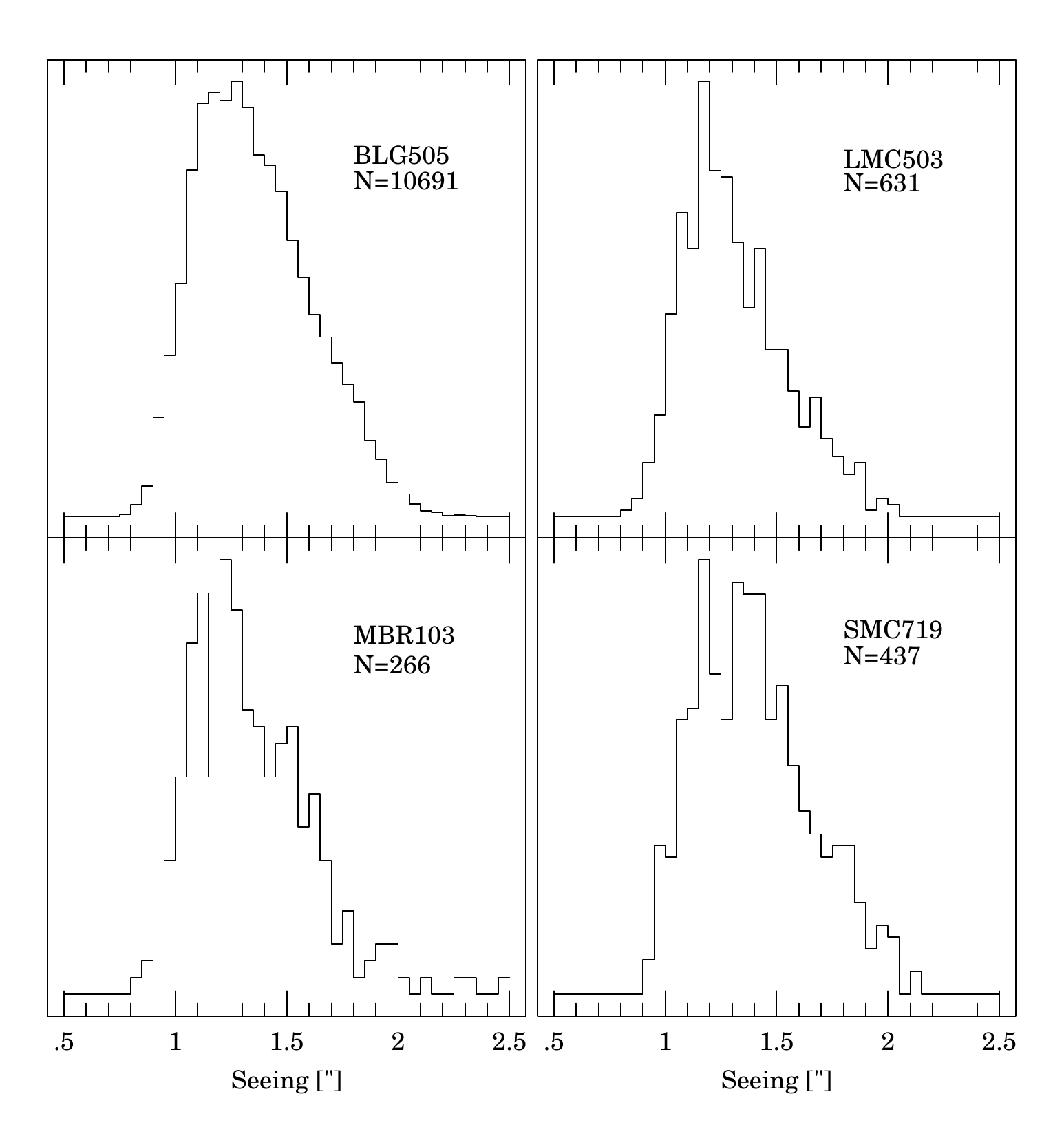}}
\vskip-3mm
\FigCap{Distribution of measurements of the PSF FWHM in the images of
the OGLE-IV fields (Warsaw telescope seeing): one of the highest stellar
density in the Galactic bulge, BLG505, central crowded fields of the
Magellanic Clouds, LMC503 and SMC719, very empty field in the Magellanic
Bridge, MBR103. N denotes number of observations. Sharp cutoff near
$0\zdot\arcs95$ is caused by the telescope optics, dome seeing and the
need to optimally focus the telescope across very large field of view.
Usually the science data are not acquired when the seeing exceeds
$2\arcs$.}
\end{figure}

The rich collection of images gathered during the OGLE-IV phase
provides precise data for testing long term image quality of the
Warsaw telescope at the Las Campanas site. Fig.~14 shows the
distribution of seeing (actually FWHM of stellar PSF) measured in one
of the most stellar dense fields in the sky -- the Galactic bulge
field BLG505 as well as crowded central fields from the Magellanic
Clouds, SMC719 and LMC503, and empty field from the Magellanic Bridge,
MBR103. The distribution of seeing is clearly non-symmetric with a
sharp cutoff at $0\zdot\arcs95$ resulting from the dome seeing,
telescope classical optics and a requirement of optimal focusing of
the entire CCD mosaic. It is also slightly shifted toward larger
seeing values for the Magellanic System fields which are lower in the
sky. The median seeing ranges from $1\zdot\arcs25$ to $1\zdot\arcs35$.

The OGLE-IV survey consists of a series of large-scale wide-field
surveys. They are divided into two categories.

\Subsection{Short-Term Large-Scale Sky Surveys}
Short-term OGLE-IV surveys are designed to be conducted for a limited
amount of time, typically for one observing season. The best example of
this type of survey was the OGLE Carnegie Kuiper Belt Survey (OCKS,
Sheppard \etal 2011). During OCKS about 2200 square degrees of the so far
unexplored sky located about 20 degrees south from the southern sky
ecliptic and near the Galactic plane (light gray contours in Fig.~13)
were monitored for new solar system bodies -- large size Kuiper Belt
objects, candidates for dwarf planets. For the detection of moving objects
located at about 30--50~a.u. three images separated by about 2--3 hours had
to be taken for each pointing. A customary software based on the image
subtraction technique, routinely used by the OGLE survey for data
reduction, was written for the detection of moving objects.

The OCKS was conducted for one season (2010). About 5100 images in 1550
pointings were collected during this project. Altogether 14 new large
Kuiper Belt objects were detected (Sheppard \etal 2011). One of the best
candidates, 2010~EK139 has been followed up by the Herschel satellite
for albedo determination what allowed the determination of its size --
equal to 470~km (P{\'a}l \etal 2012). 2010~EK139, being a scattered
disk object, is one of the best candidates for a dwarf planet.  

Another type of short-time surveys conducted from time to time by OGLE are
observations of large areas of the sky for the detection of stellar
streams, etc.

\begin{landscape}
\begin{figure}[p]
\vglue-7mm{\centerline{\includegraphics[width=19cm]{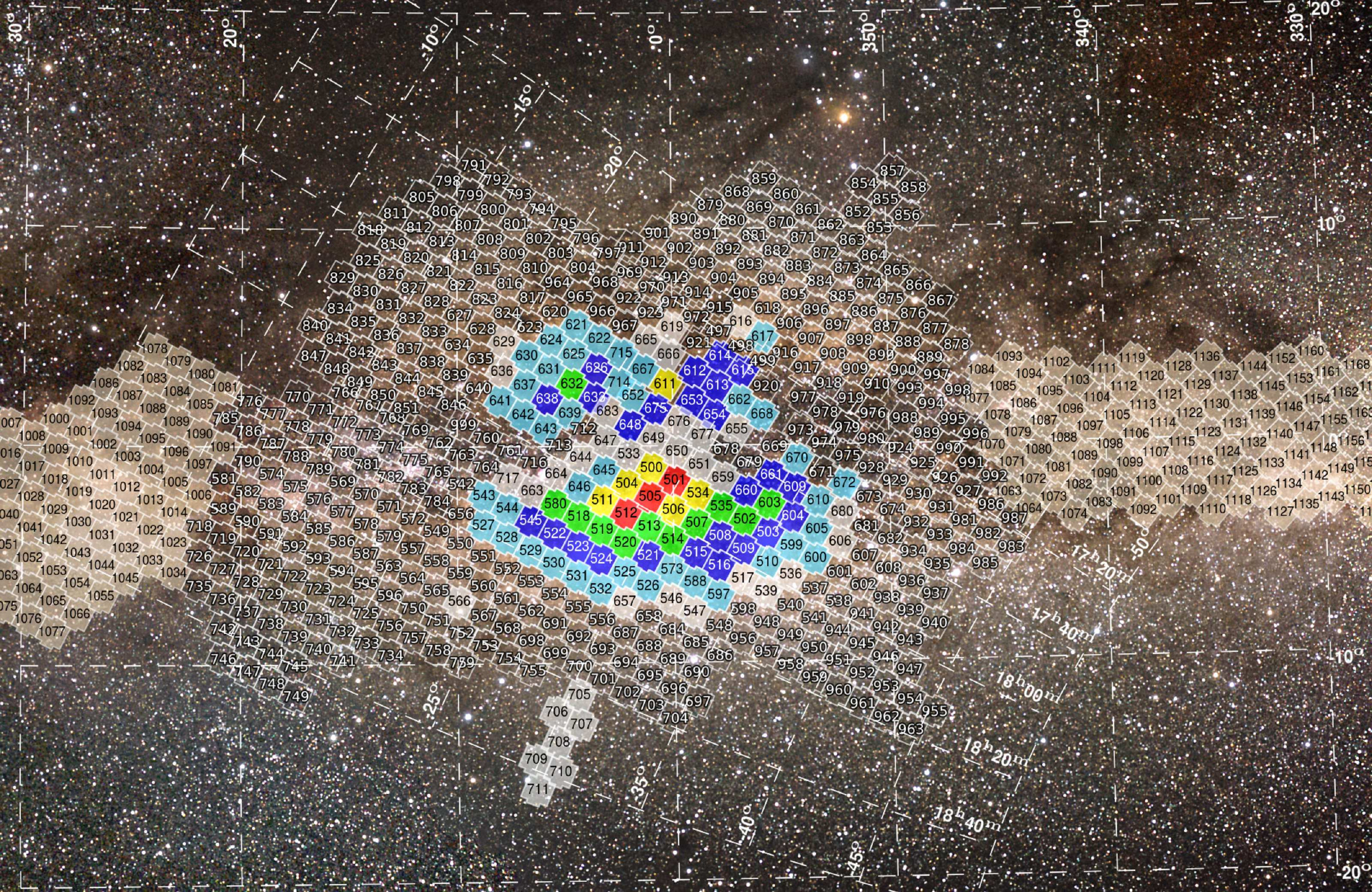}}}
\vskip4pt
\FigCap{OGLE-IV Galactic center fields. Cadence of observations of the
inner Galactic bulge microlensing fields is color-coded. On average: red
fields -- one observation every 19 minutes throughout the visibility of
the field ($z<62$~deg), yellow fields -- one observation per hour,
green fields -- 2--3 observations per night, dark blue fields -- one
observation per night, cyan fields -- one observation per two nights.
Silver color marks additional fields which were regularly observed in
the years 2010--2013. Transparent fields in the outer Galactic bulge and
beige fields of the Galactic disk are monitored during the OGLE Galaxy
Variability Survey (see also Fig.~21).  Credit: Jan Skowron, background
photograph: Krzysztof Ulaczyk.}
\end{figure}
\end{landscape}

\Subsection{Long-Term Large-Scale Sky Surveys}
Long-term OGLE-IV surveys constitute a category of surveys lasting for
extended amount of time, several observing seasons. The main
scientific aim of these surveys is to provide the sky variability
information in the selected OGLE-IV lines-of-sight. However, very
precise OGLE photometry can be simultaneously used for mapping
non-variable objects for many scientific projects related, for example
with the structure studies of observed targets, etc.

\vspace{7pt}
{\it Galactic Bulge}
\vspace{3pt}

The main OGLE-IV long-term surveys conducted from the beginning of the
OGLE-IV phase include the second generation microlensing survey of the
Galactic bulge. It monitors regularly over 130 square degrees in the
Galactic center -- see the sky map in Fig.~15. The Galactic bulge fields
are observed with different cadence which is color-coded in Fig.~15. It
depends on the frequency of microlensing events observed in these
fields. Fig.~16 shows the histogram of the difference of epochs between
consecutive observations for typical fields of different categories
based on observations collected in the years 2010--2014. Histograms of
every category are color-coded as in Fig.~15 and normalized to the
highest peak.

\begin{figure}[ht]
\centerline{\includegraphics[width=11cm]{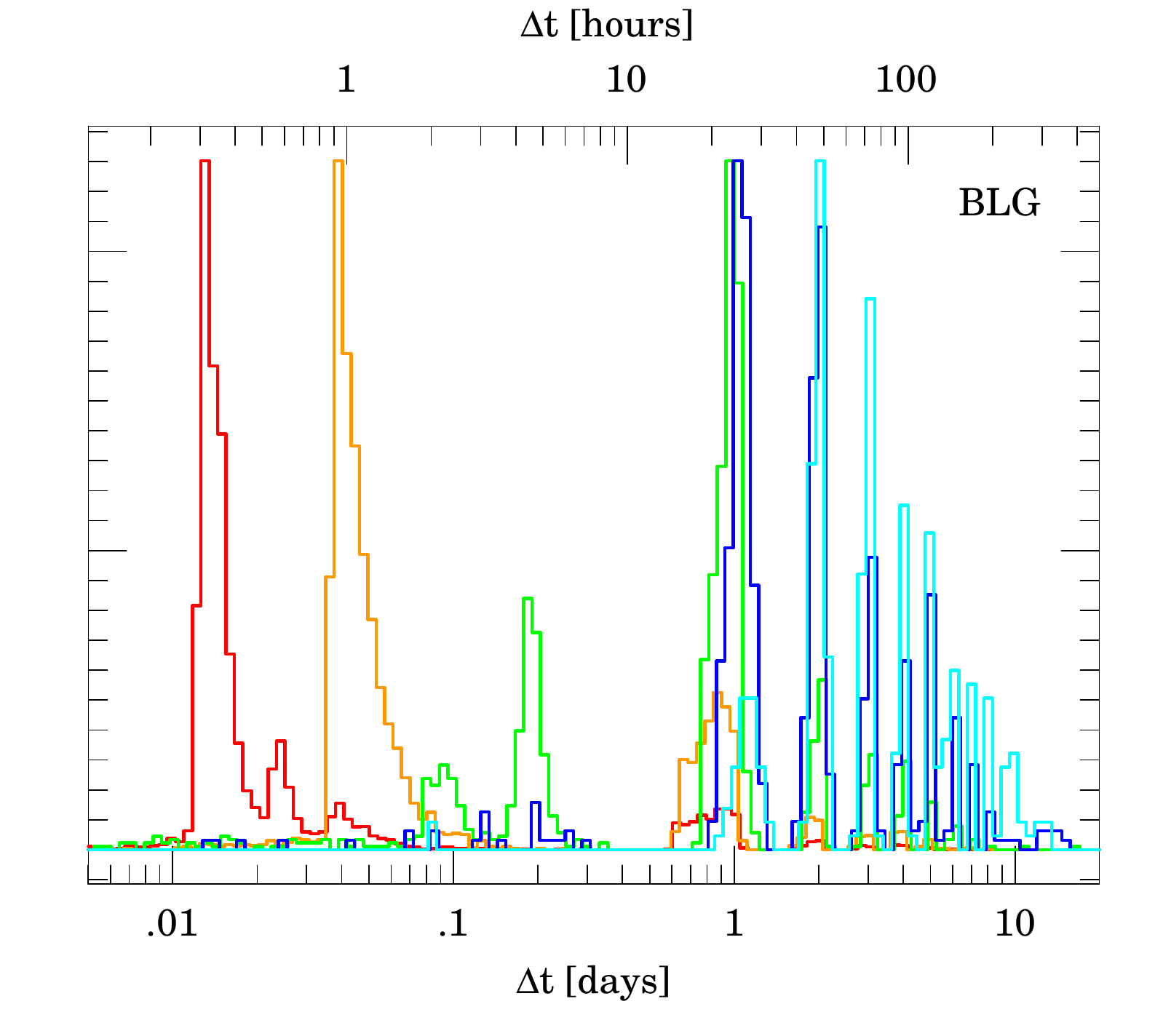}}
\FigCap{Cadence of observations of the second generation microlensing
survey fields: histograms of the time difference between consecutive
observations. The histograms are normalized to the highest peak and
color-coded as in Fig.~15.}
\end{figure}
The first two categories of fields (red and yellow in Figs.~15 and 16)
cover 4.5 and 8.5 square degrees in the sky and are observed with the
cadence of 19 and 60~minutes (their main peaks in Fig.~16), respectively.
Lower peaks in Fig.~16 correspond to the typical frequency of observations
during early (February-March) and late (September-October) parts of the
observing season. These are the main fields of the second generation
microlensing survey focused on the detection of a large number of
gravitational microlensing events for studying exoplanets. The microlensing
technique of exoplanet detection proved to be very important tool for
exoplanet studies (\eg Poleski \etal 2014ab, Gould \etal 2014, Skowron 
\etal 2015). 

\begin{figure}[b]
\centerline{\includegraphics[width=11.3cm]{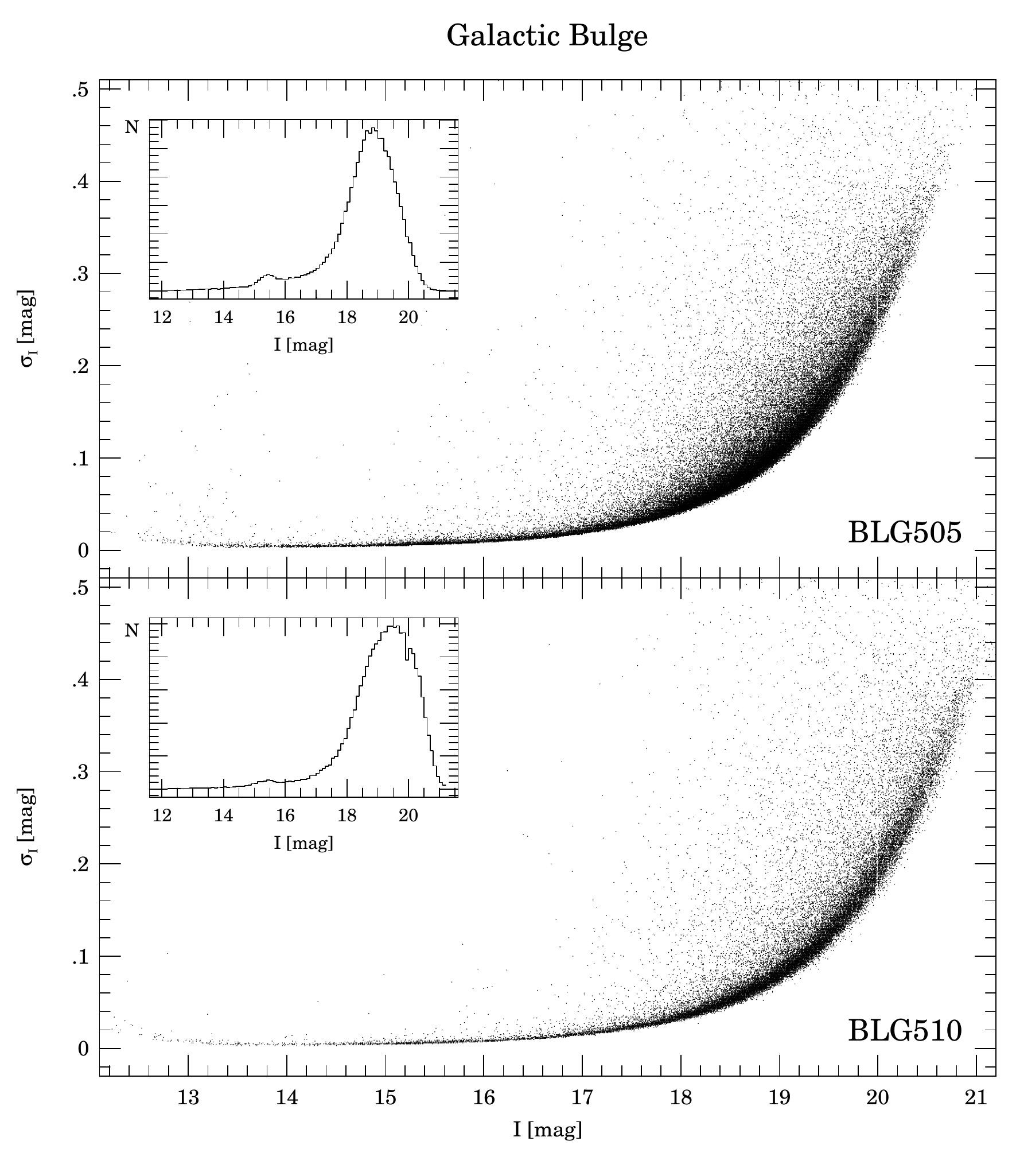}}
\FigCap{Accuracy of the OGLE-IV {\it I}-band photometry: standard
deviation of all observations of a star collected in the years 2010--2014
as a function of its {\it I}-band magnitude for the most crowded highest
cadence field BLG505 and somewhat lower stellar density field BLG510. Only
25\% stars are plotted, for clarity. Insets present histograms of the
number of detected stars as a function of magnitude illustrating
completeness of the OGLE-IV photometry.}
\end{figure}
The fields of the remaining categories are monitored with lower
cadence, typically: green fields 1--3 times per night, blue -- once
per night, cyan once per two nights (during the main part of the
observing season).

The total number of stars monitored during the second generation
microlensing survey exceeds 400 million. Typical number of collected
science images of the Galactic bulge for this survey reaches about 23
thousand per year ($\approx12.5$ TBytes/season). The total number of
collected epochs in the {\it I}-band during the seasons 2010--2014
varies from about 10\,700 to 200 per field for the highest (red) and
lowest (cyan) cadence fields, respectively. The fields centered on the
Sagittarius Dwarf galaxy (BLG705--BLG711) were observed over 150 times
and the highly obscured fields at the Galactic plane about 65--100
times. Additionally 3670 {\it V}-band images were collected (6--128 per
field).

Fig.~17 shows the accuracy of the {\it I}-band photometry and
completeness of the survey in one of the highest stellar density fields
-- BLG505 and typical Galactic bulge field of lower density -- BLG510.
The OGLE-IV photometry covers the range of $12<I<20.5$~mag and
$12<I<21$~mag in the BLG505 and BLG510 fields, respectively. However,
due to high stellar crowding it is complete to $I\approx18.5$~mag and
$I\approx19$~mag in these fields.

Beside studies of exoplanets with microlensing, the photometric data of the
second generation microlensing survey are ideal observational material for
investigating all variety of variables in the Galactic bulge (\eg RR~Lyr
stars, Soszyñski \etal 2014, Pietrukowicz \etal 2015). Data for
non-variable stars are perfect tool for studying, for example, the Galactic
structure (Nataf \etal 2010, 2015).

\vspace{11pt}
{\it Magellanic System}
\vspace{7pt}

The second main long-term OGLE-IV survey conducted mostly during
complementary part of the year is the monitoring of about 650 square
degrees of the Magellanic System. This huge survey provides variability
data for the Magellanic Cloud objects, as well as for the Magellanic Bridge
(Fig.~18). Non-variable objects detected during this survey for the first
time allowed the precise mapping of the stellar populations in the
Magellanic Bridge (Skowron \etal 2014). The Magellanic System survey also
provides data for the OGLE-IV extragalactic projects -- OGLE-IV SN survey
(Wyrzykowski \etal 2014) or The Magellanic Quasars Survey (Koz³owski \etal
2013).

Fig.~19 shows the distribution of time difference between consecutive
epochs of the selected OGLE-IV fields in the Magellanic System fields based
on observations collected in the years 2010--2014. This plot -- similar to
Fig.~16 for the Galactic bulge -- indicates that the typical cadence of the
central fields in the Large and Small Magellanic Clouds was about 1--3~days
and 2--3 days for the main Magellanic Bridge fields. Halo fields of both
Clouds (not shown in Fig.~19) were observed with a somewhat lower cadence
of 3--5~days.

\begin{landscape}
\begin{figure}[ht]
\vglue-5mm{\centerline{\includegraphics[width=19cm]{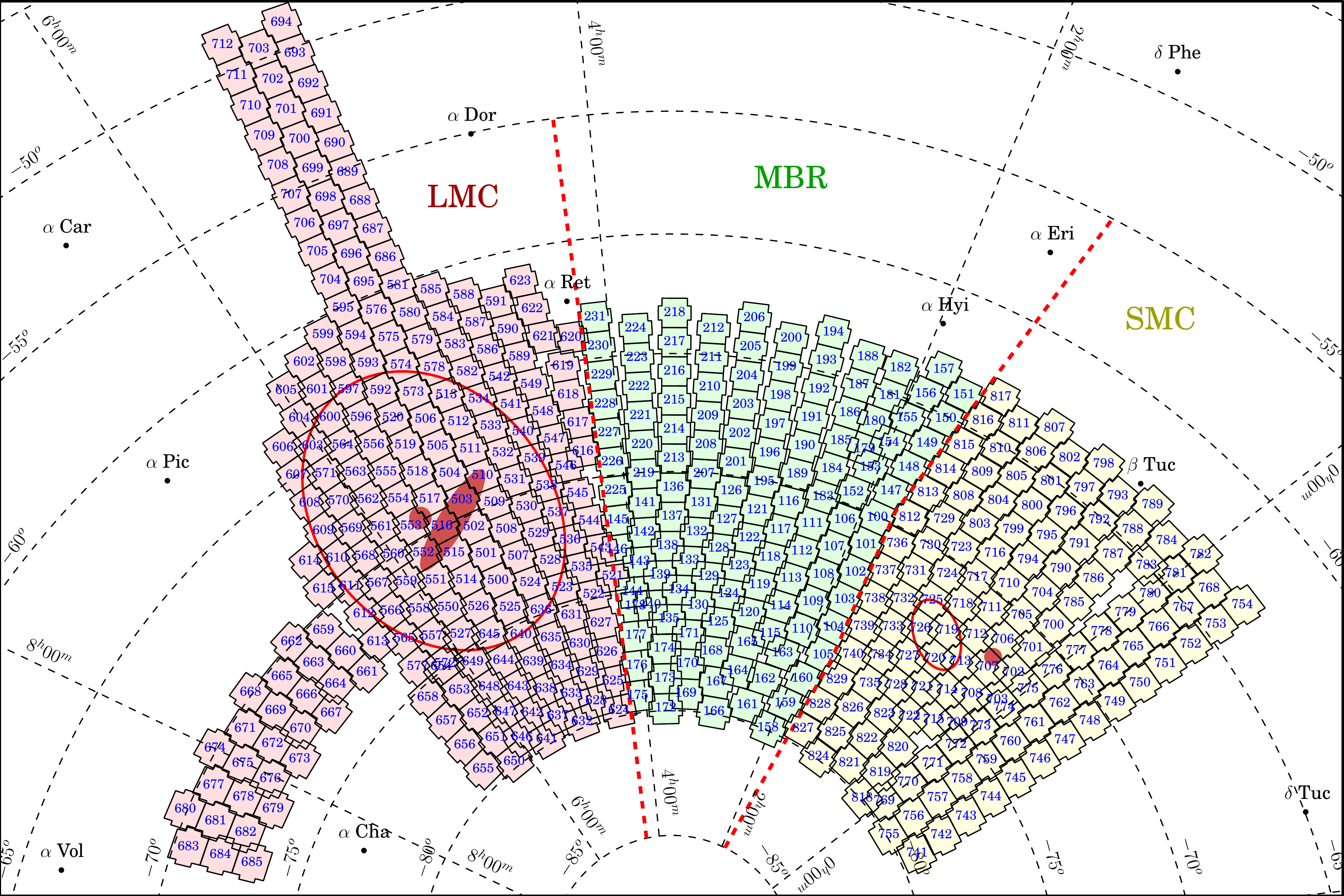}}}
\vspace{2mm}
\FigCap{OGLE-IV coverage of the Magellanic System. Big, open ellipses
mark LMC disk and the core of the SMC. Filled ellipses mark LMC bar,
Tarantula star forming region in the LMC and 47 Tuc near the SMC. Survey
fields are artificially divided into three sky regions: LMC, MBR and
SMC. Fields are named with consecutive numbers in the given region. Each
field is surveyed in the {\it I}- and {\it V}-bands (see text for
cadence and number of science frames already taken). Credit: Jan
Skowron.}
\end{figure}
\end{landscape}
\begin{figure}[ht]
\vglue-2mm
\centerline{\includegraphics[width=10.5cm]{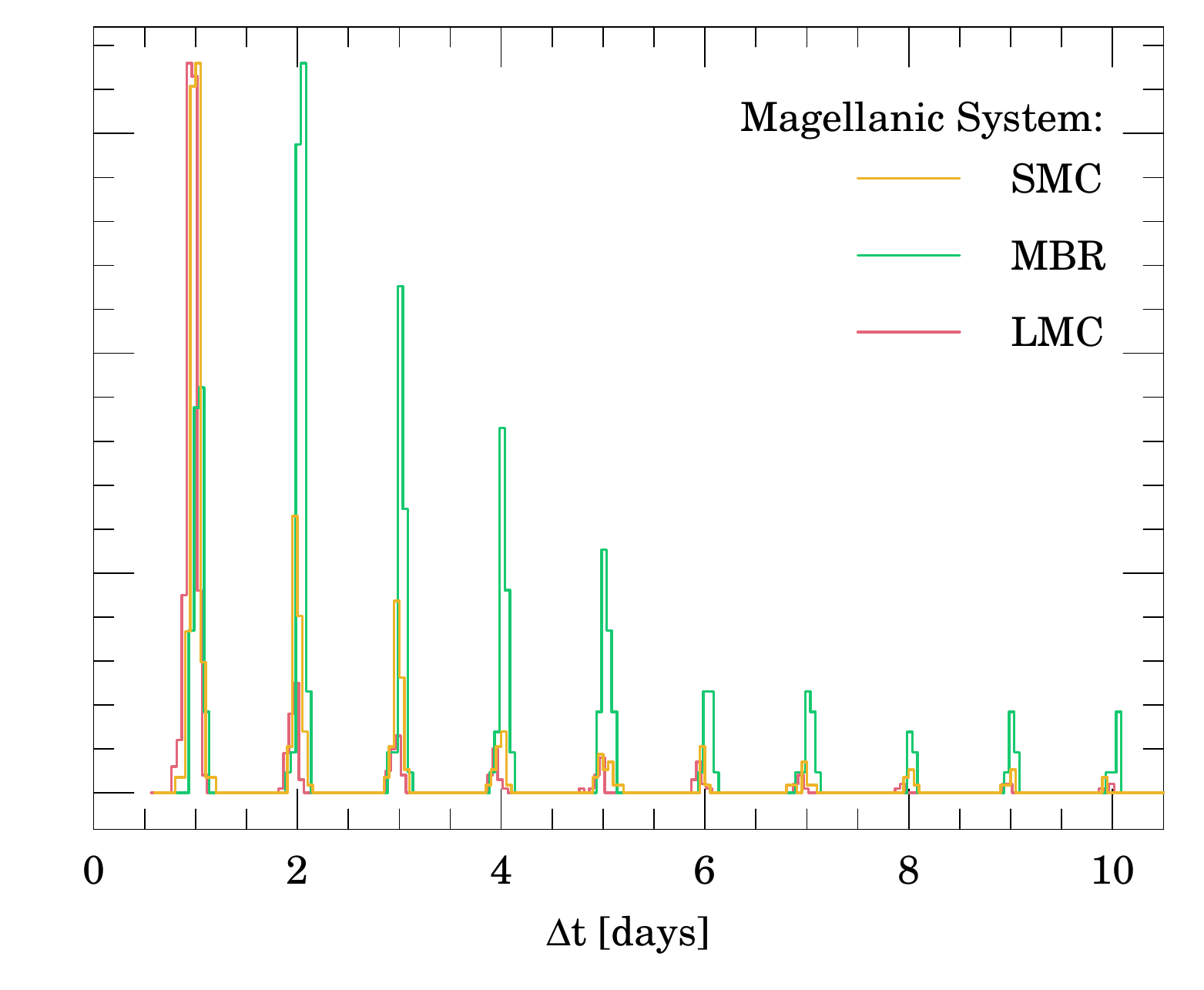}}
\FigCap{Cadence of observations of the main Magellanic System fields:
histograms of the time difference of the consecutive observations,
normalized to the highest peak.}
\end{figure}
\begin{figure}[ht]
\centerline{\includegraphics[width=11.4cm]{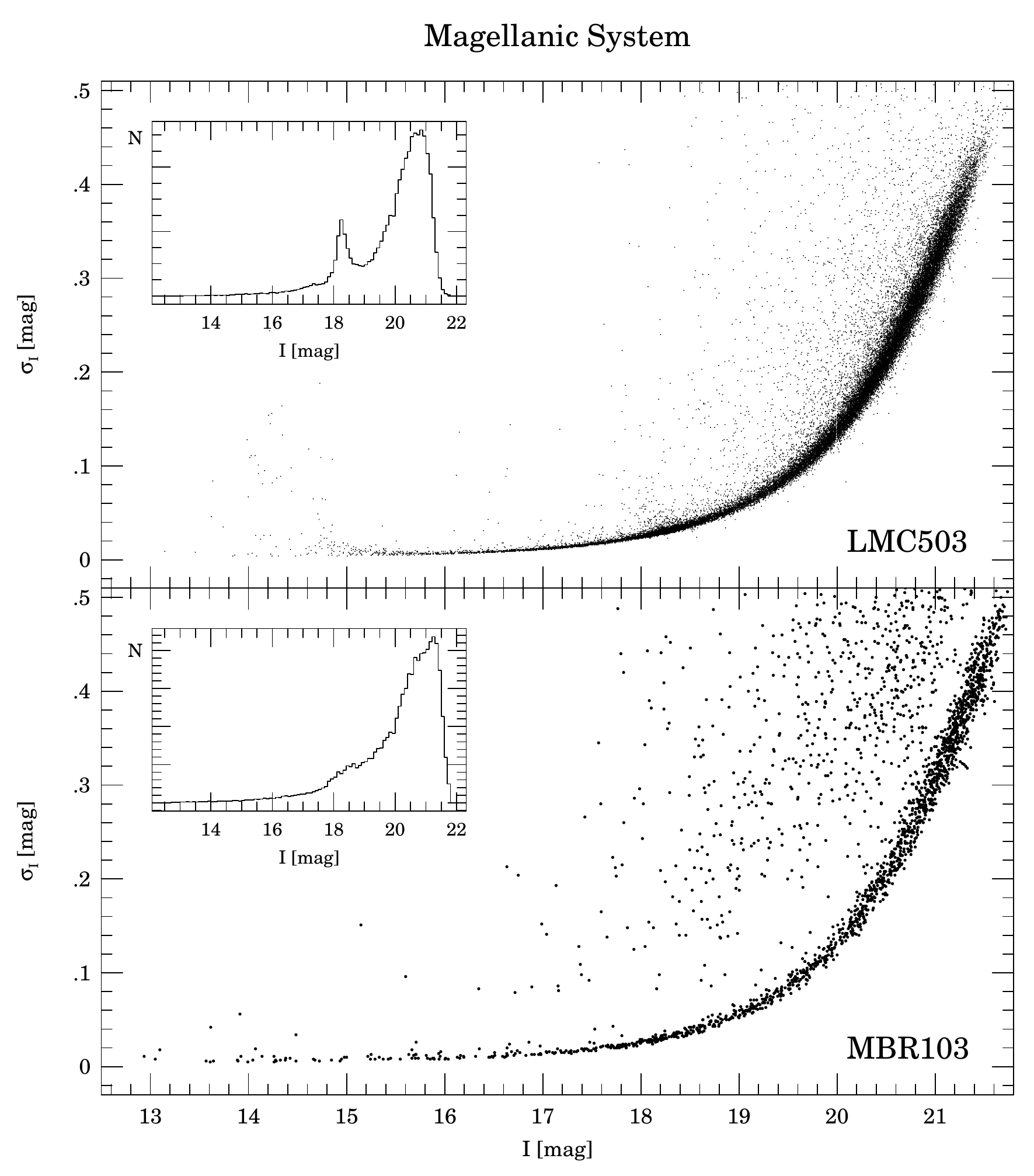}}
\FigCap{Accuracy of the OGLE-IV {\it I}-band photometry of the typical
Magellanic System fields: high stellar density field LMC503 from the
center of the LMC and very empty Magellanic Bridge field MBR103. Only
33\% stars from the LMC503 field were plotted, for clarity. Insets
show completeness of photometry -- histograms of stellar counts as a
function of magnitude. MC red clump giants are responsible for the
excess of stars at $18.0<I<18.5$~mag.}
\end{figure}
\noindent
The total number of monitored stars in the Magellanic System exceeds
80 million (62, 13, 6 million in the LMC, SMC and MBR, respectively).
Typically 23 thousand science images of these fields are collected
every observing season (lasting approximately from June to May next
year). This is $\approx12.5$~TBytes of raw data per season. The total
number of {\it I}-band observations (for variability studies) per
field collected during the seasons 2010/2011--2013/2014 varies from
300--700, 300--500 and 270--310 epochs in the central fields of the
LMC, SMC and main part of MBR, respectively, For the halo fields the
number of collected observations was in the range of 82--150 epochs
per field. In the {\it V}-band the number of collected science images
was about 8700: 10--230 per field in the main parts of the Magellanic
Clouds and Bridge and 3--10 in the halo fields.

OGLE-IV {\it I}-band photometry covers the range of about $13<I<21.7$~mag
in the Magellanic System fields. Fig.~20 presents the accuracy of
photometry and its completeness in the central LMC field of high stellar
density, LMC503, and typical empty field in the Magellanic Bridge --
MBR103. Full completeness of photometry reaches $I\approx20.5$~mag in the
very dense stellar fields like LMC503 and $I\approx21.2$~mag in the empty
fields like MBR103. Obviously, this is a result of stellar crowding and
higher background in the former fields.

\vspace{7pt}
{\it OGLE Galaxy Variability Survey}
\vspace{3pt}

\begin{figure}[p]
\centerline{\includegraphics[width=10.5cm]{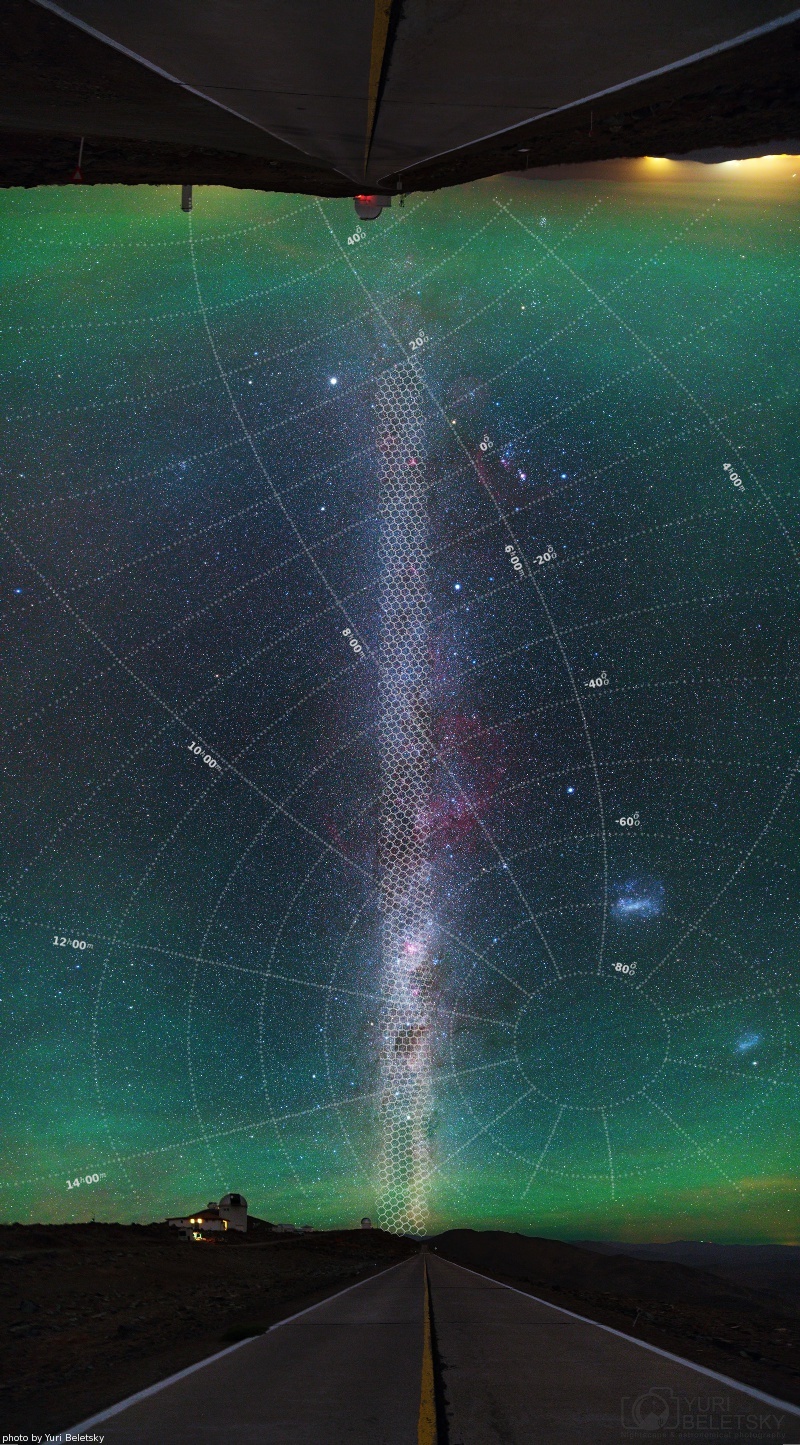}}
\vspace{4mm}
\FigCap{Part of the sky over the Las Campanas Observatory observed
during the OGLE Galaxy Variability Survey. Warsaw telescope is visible
on the left. Credit: Jan Skowron. Background photograph of the Milky Way
from Las Campanas Observatory: Yuri Beletsky.}
\end{figure}
In 2013 another long-term sky variability survey was started by
OGLE-IV.  This is the largest ever, long-term photometric survey of
the Galaxy -- OGLE Galaxy Variability Survey. By regular monitoring
(1--2~days cadence for the first $\approx100$ epochs and lower after
that) of over 2000 square degrees -- around the Galactic plane and
extended area around the outer Galactic bulge -- this survey will
provide unique variability census of the Milky Way as seen from Las
Campanas Observatory (see Figs.~15 and 21), up to
$I\approx18.5$~mag. These data will be extremely valuable for studying
the Galaxy structure and variable objects properties. A complementary,
much deeper survey started yet in 2010 will map all stellar objects
from these lines-of-sight to $I\approx22$~mag. In the observing
seasons 2010--2014 about 56\,600 images in the {\it I}-band and 4400
{\it V}-band frames were collected by OGLE-IV survey for this project
(34 TBytes of raw data).

\Subsection{Prospects}
OGLE-IV survey has been successfully run for the last five years. After
the initial phase of the implementation of hardware and basic software,
implementation of real time reductions and real time systems, the
OGLE-IV survey reached its expected performance and entered the phase of
regular observing and real time data reductions. Tens of scientific
projects has already been completed and their results published.

Information on the current status of the OGLE project and its
discoveries can be obtained at:

\centerline{{\it http://ogle.astrouw.edu.pl}}
\vspace*{2pt}
\noindent
This is also a starting point for data-mining of the huge OGLE Internet
Archive.

During the next years the OGLE-IV survey will continue monitoring of the
most interesting regions of the sky focusing on the densest stellar
areas, usually omitted by other wide-field sky surveys. OGLE-IV will
provide new collections of variable objects and photometric databases
which will be a gold mine for hundreds of new scientific projects for
years to come.

\vspace*{-5pt}
\Acknow{The OGLE project has received funding from the European Research
Council under the European Community's Seventh Framework Programme
(FP7/2007-2013)/ERC grant agreement no. 246678 to AU.

We would like to express our greatest thanks to the Las Campanas
Observatory personnel for generous help during preparatory phase of the
OGLE-IV hardware assembling, during the instrument commissioning and for
continuous support for the OGLE project during its entire presence at
the Las Campanas Observatory. In particular we thank Don Oscar Duhalde
for his invaluable help in many critical situations.

We thank staff of Zak³ady Mechaniczne Kazimieruk Sp.~z o.o., in
particular Mr.\ Krystian Wojtaszak, for very helpful attitude during the
manufacturing of the OGLE-IV camera mechanical parts.

Finally, we would like to thank all OGLE team members who actually
contribute actively to the OGLE scientific outcome and successes. We are
very grateful to Drs.\ Szymon Koz³owski, Jan Skowron, Pawe³
Pietrukowicz, Radek Poleski and Igor Soszyñski for comments on the
manuscript and preparation of selected figures. We also thank Drs.\ Yuri
Beletsky and Krzysztof Ulaczyk for the beautiful photographs of the sky
used in Figures in this paper.}

\end{document}